\documentclass[twocolumn,english,prb,showpacs]{revtex4-1}

\usepackage{float}
\usepackage[dvips,final]{graphicx}
\usepackage{color}
\usepackage{amsmath}

\usepackage{ulem}
\usepackage{graphicx, subfigure}
\usepackage{pslatex}
\usepackage{relsize}
\usepackage{bm}
\usepackage{verbatim}

\usepackage{appendix}

\usepackage{booktabs}
\usepackage{xcolor}
\usepackage{soul}

\usepackage{natbib}

\begin{document}
\title{Atomistic simulations of magnetoelastic effects on sound velocity}

\author{P. Nieves$^{1}$}
\email{Corresponding author: pablo.nieves.cordones@vsb.cz}
\author{J. Tranchida$^2$}
\author{S. Nikolov$^3$}
\author{A. Fraile$^{4}$}

\author{D. Legut$^{1}$}

\affiliation{$^1$ IT4Innovations, V\v{S}B - Technical University of Ostrava, 17. listopadu 2172/15, 70800 Ostrava-Poruba, Czech Republic}
\affiliation{$^2$ CEA, DES/IRESNE/DEC, 13018 Saint Paul Lès Durance, France}
\affiliation{$^3$ Computational Multiscale Department, Sandia National Laboratories, P.O. Box 5800, MS 1322, 87185 Albuquerque, NM, United States}
\affiliation{$^4$ Nuclear Futures Institute, Bangor University, Bangor, LL57 1UT, United Kingdom of Great Britain and Northern Ireland}

\date{\today}

\begin{abstract}
In this work, we leverage atomistic spin-lattice simulations to examine how magnetic interactions impact the propagation of sound waves through a ferromagnetic material. To achieve this, we characterize the sound wave velocity in BCC iron, a prototypical ferromagnetic material, using three different approaches that are based on the oscillations of kinetic energy, finite-displacement derived forces, and corrections to the elastic constants, respectively. Successfully applying these methods within the spin-lattice framework, we find good agreement with the Simon effect including high order terms. In analogy to experiments, morphic coefficients associated with the transverse and longitudinal waves propagating along the [001] direction are extracted from fits to the fractional change in velocity data. The present efforts represent an advancement in magnetoelastic modelling capabilities which can expedite the design of future magneto-acoustic devices.


\end{abstract}
\pagebreak
\maketitle

\section{Introduction}

Magnetoelastic (MEL) interactions are responsible for many interesting phenomena in magnetic materials\cite{booktremolet} such as Joule magnetostriction\cite{Joule}, the Wiedemann effect\cite{MALYUGIN1991193}, the Villari effect\cite{booktremolet}, the Matteucci effect\cite{Mateucci}, anomalous thermal expansion\cite{WASSERMAN1990237}, and many others\cite{booktremolet,Benito2007,Adroja2012}. MEL coupling also leads to  complex effects on sound velocity  that took about four decades, from the first works by Fuchs\cite{Fuchs} in 1936 and Mueller\cite{Mueller} in 1940 until the comprehensive study by Rouchy et al.\cite{Rouchy1979} in 1979, to fully understand  \cite{ROUCHY198069,booktremolet}. Four main magnetic effects on sound velocity have been identified: (i) isotropic exchange effects\cite{Fuchs,Isenberg}, (ii) anisotropic morphic effects\cite{Mason1951b,Eastman,Rouchy1979}, (iii) field dependent effects (the Simon effect) \cite{Simon+1958+84+89,Sato1958,sakurai} and (iv) rotational-magnetostrictive effects \cite{Rouchy1979,ROUCHY198069}. J. Rouchy and E. du Tremolet de Lacheisserie provided a detailed theoretical derivation of these four effects for cubic crystals by expanding the internal energy as a series of
the Lagrangian tensor components, as well as symmetrical and
antisymmetrical components of the homogeneous
strains \cite{Rouchy1979}. S. Rinaldi and G. Turilli showed that MEL effects on sound velocity can  be equivalently taken into account as corrections to the elastic constants \cite{Rinaldi1985}. These MEL effects can be large, and have been experimentally observed in many materials through the dependency of ultrasonic sound wave velocity on the intensity and direction of an applied magnetic field\cite{ROUCHY198069,DUTREMOLETDELACHEISSERIE198277,ROUCHY198159,sakurai,ALERS195921,Dietz1976,booktremolet,kingner,SEAVEY1972219}. Novel magneto-acoustic phenomena have been discovered in recent years, like acoustic spin pumping\cite{Uchida2011} and magnetization switching induced by sound waves\cite{Camara,Kovalenko,Thevenard2013,Thevenard2016}, with potential technological applications in spintronics and magnetic recording\cite{Weiyang}.

Until recently, atomistic simulations of magnetic effects on sound waves were quite challenging. This phenomenon involves a coupled dynamics of magnetic moments and atoms, so that it is not possible to use only standard atomistic spin dynamics (SD) or molecular dynamics (MD) since the motion of atoms or spins are neglected, respectively \cite{Evans_2014,Eriksson_2017,thompson2022lammps}. 
An alternative approach could be to combine spin-polarized ab-inito molecular dynamics and SD \cite{Stockem}. This strategy is very accurate but unfortunately is quite demanding computationally, so that it might not a be a convenient method  due to the large atomic supercells required to study sound waves. To overcome this limitation, one could use classical “spin-lattice dynamics”\cite{Ma2008,Beaujouan2012,MA2016350,Wu2018,Perera2017,TRANCHIDA2018406} (SD-MD) that combines SD and MD, and enables the simulation of both large system sizes and time scales\cite{TRANCHIDA2018406,MA2016350}. Recent advances in atomistic models based on spin-lattice simulations offer the possibility of studying MEL phenomena computationally \cite{TRANCHIDA2018406,Nikolov2021}. In our previous work, we presented a methodology based on the N\'{e}el model to build a classical spin-lattice Hamiltonian for cubic crystals capable of describing  magnetic properties induced by the spin-orbit coupling (SOC) like magnetocrystalline anisotropy (MCA) and anisotropic magnetostriction, as well as exchange magnetostriction \cite{nieves2021spinlattice_prb}. 
Here, we probe the range of applicability of these models by simulating the MEL effects on sound velocity.
Such kind of atomistic simulations may be useful to clarify and further understand the physics of this complex phenomenon, as well as speed-up the design of possible novel technological applications based on these effects.

\section{Methodology}
\label{section:method}

\subsection{Spin-Lattice Hamiltonian}
\label{section:SL}

For the atomistic spin-lattice simulations  we consider the following Hamiltonian 
\begin{equation}
\begin{aligned}
\mathcal{H}_{sl}(\boldsymbol{r},\boldsymbol{p},\boldsymbol{s}) & =  \mathcal{H}_{mag}(\boldsymbol{r},\boldsymbol{s})+\sum_{i=1}^N\frac{\boldsymbol{p}_i}{2m_i}+\sum_{i,j=1}^N\mathcal{V}(r_{ij}),
\label{eq:Ham_tot}
\end{aligned}
\end{equation}
where $\boldsymbol{r}_i$, $\boldsymbol{p}_i$, $\boldsymbol{s}_i$, and $m_i$ stand for the position, momentum, normalized magnetic moment and mass for each atom $i$ in the system, respectively, $\mathcal{V}(r_{ij})=\mathcal{V}(\vert \boldsymbol{r}_i-\boldsymbol{r}_j\vert)$ is the interatomic potential energy and $N$ is the total number of atoms in the system with total volume $V$. Here, we include the following interactions in the magnetic energy
\begin{equation}
\begin{aligned}
\mathcal{H}_{mag}(\boldsymbol{r},\boldsymbol{s}) & =  -\mu_{0}\sum_{i=1}^N\mu_i\boldsymbol{H}\cdot\boldsymbol{s}_i -\frac{1}{2}\sum_{i,j=1,i\neq j}^N J(r_{ij})\boldsymbol{s}_i\cdot\boldsymbol{s}_j \\
& + \mathcal{H}_{N\acute{e}el}(\boldsymbol{r},\boldsymbol{s}),
\label{eq:Ham_mag}
\end{aligned}
\end{equation}
where $\mu_{i}$ is the atomic magnetic moment, $\mu_0$ is the vacuum permeability, $\boldsymbol{H}$ is the external magnetic field, $J(r_{ij})$ is the exchange parameter. The term $\mathcal{H}_{N\acute{e}el}$ is the N\'{e}el interaction
\begin{equation}
\begin{aligned}
\mathcal{H}_{N\acute{e}el} & =   -\frac{1}{2}\sum_{i,j=1}^{N} \lbrace g(r_{ij}) + l_1(r_{ij})\left[ (\boldsymbol{e}_{ij}\cdot\boldsymbol{s}_i)(\boldsymbol{e}_{ij}\cdot\boldsymbol{s}_j)-\frac{\boldsymbol{s}_i\cdot\boldsymbol{s}_j}{3}\right]  \\
& + q_1(r_{ij})\left[ (\boldsymbol{e}_{ij}\cdot\boldsymbol{s}_i)^2-\frac{\boldsymbol{s}_i\cdot\boldsymbol{s}_j}{3}\right]\left[ (\boldsymbol{e}_{ij}\cdot\boldsymbol{s}_j)^2-\frac{\boldsymbol{s}_i\cdot\boldsymbol{s}_j}{3}\right]   \\
& + q_2(r_{ij})\left[ (\boldsymbol{e}_{ij}\cdot\boldsymbol{s}_i)(\boldsymbol{e}_{ij}\cdot\boldsymbol{s}_j)^3+(\boldsymbol{e}_{ij}\cdot\boldsymbol{s}_j)(\boldsymbol{e}_{ij}\cdot\boldsymbol{s}_i)^3\right]  \rbrace,
\label{eq:Neel_energy}
\end{aligned}
\end{equation}
where $\boldsymbol{e}_{ij}=\boldsymbol{r}_{ij}/r_{ij}$, and 
\begin{equation}
\begin{aligned}
l_1(r_{ij}) & = l(r_{ij})+\frac{12}{35}q(r_{ij}), \\
q_1(r_{ij}) & = \frac{9}{5}q(r_{ij}), \\
q_2(r_{ij}) & = -\frac{2}{5}q(r_{ij}). \\
\end{aligned}
\end{equation}
In the case of a collinear state ($\boldsymbol{s}_i\parallel\boldsymbol{s}_j$), the Eq. \ref{eq:Neel_energy} is reduced to
\begin{equation}
\begin{aligned}
\mathcal{H}_{N\acute{e}el} & =   - \frac{1}{2}\sum_{i,j=1}^{N} \lbrace g(r_{ij}) +l(r_{ij}) \left(\cos^2\psi_{ij} -\frac{1}{3}\right) \\
& + q(r_{ij}) \left( \cos^4\psi_{ij} -\frac{6}{7}\cos^2\psi_{ij} + \frac{3}{35}\right) \rbrace
\label{eq:Neel_energy_coll}
\end{aligned}
\end{equation}
where $\cos\psi_{ij}=\boldsymbol{e}_{ij}\cdot\boldsymbol{s}_i$.  The N\'{e}el energy reproduces the correct symmetry of MCA and MEL energy\cite{Skomski}. The quantity $g(r_{ij})$ is set to
\begin{equation}
\begin{aligned}
g(r_{ij})=J(r_{ij}),
\label{eq:g}
\end{aligned}
\end{equation}
in order to offset the exchange interaction, as detailed in Ma et al. \cite{Ma2008}. This offset of the exchange energy does not affect the precession dynamics of the spins. However, it allows to offset the corresponding mechanical forces. Without this additional term, the forces and the pressure generated by the magnetic Hamiltonian are not zero at its energy ground state (corresponding to the ferromagnetic state). By doing so, the function $g(r_{ij})$ can also be related to the volume magnetostriction $\omega_s$ induced by the exchange interactions\cite{Chika,nieves2021spinlattice_prb}. On the other hand, the dipole ($l(r_{ij})$) and quadrupole ($q(r_{ij})$) terms can describe the effects induced by SOC like the anisotropic magnetostriction and MCA, respectively \cite{Chika,nieves2021spinlattice_prb}.  In this work, the spatial dependence of $J(r_{ij})$, $l(r_{ij})$ and $q(r_{ij})$ is described using the Bethe-Slater curve, as implemented in the SPIN package of LAMMPS \cite{TRANCHIDA2018406}
\begin{equation}
\begin{aligned}
J(r_{ij}) & =  4\alpha_J \left(\frac{r_{ij}}{\delta_J}\right)^2 \left[1-\gamma_J\left(\frac{r_{ij}}{\delta_J}\right)^2 e^{-\left(\frac{r_{ij}}{\delta_J}\right)^2}\right] \Theta(R_{c,J}-r_{ij}),\\
l(r_{ij}) & =  4\alpha_l \left(\frac{r_{ij}}{\delta_l}\right)^2 \left[1-\gamma_l\left(\frac{r_{ij}}{\delta_l}\right)^2 e^{-\left(\frac{r_{ij}}{\delta_l}\right)^2}\right] \Theta(R_{c,l}-r_{ij}),\\
q(r_{ij}) & =  4\alpha_q \left(\frac{r_{ij}}{\delta_q}\right)^2 \left[1-\gamma_q\left(\frac{r_{ij}}{\delta_q}\right)^2 e^{-\left(\frac{r_{ij}}{\delta_q}\right)^2}\right] \Theta(R_{c,q}-r_{ij}),
\label{eq:BS_J_g_q}
\end{aligned}
\end{equation}
where $\Theta(R_{c,n}-r_{ij})$ is the Heaviside step function and $R_{c,n}$ ($n=J,l,q$) are the cut-off radii. The parameters $\alpha_n$, $\gamma_n$, and $\delta_n$ ($n=J,l,q$) must be determined in order to reproduce the Curie temperature ($T_C$), magnetostriction and MCA, as described in Section \ref{section:model_Fe_Ni}. 

The dynamics of spins and atoms is achieved by integrating the following equations of motion:\cite{yang1980generalizations}
\begin{equation}
\begin{aligned}
\frac{d\boldsymbol{r}_i}{dt} & =\lbrace\boldsymbol{r}_i,\mathcal{H}_{sl} \rbrace,\\
\frac{d\boldsymbol{p}_i}{dt} & =\lbrace\boldsymbol{p}_i,\mathcal{H}_{sl} \rbrace,\\
\frac{d\boldsymbol{s}_i}{dt} & =\lbrace\boldsymbol{s}_i,\mathcal{H}_{sl} \rbrace,
\label{eq:eom}
\end{aligned}
\end{equation}
where
\begin{equation}
\begin{aligned}
\lbrace F,G\rbrace = \sum_{i=1}^{N}\left(\frac{dF}{d\boldsymbol{r}_i}\cdot\frac{dG}{d\boldsymbol{p}_i}-\frac{dG}{d\boldsymbol{r}_i}\cdot\frac{dF}{d\boldsymbol{p}_i}-\frac{\boldsymbol{s}_i}{\hbar}\left[\frac{dF}{d\boldsymbol{s}_i}\times\frac{dG}{d\boldsymbol{s}_i}\right]\right).
\label{eq:eom2}
\end{aligned}
\end{equation}
In this work, we focus on MEL effects at zero-temperature, so that Langevin thermostats are not included in Eqs.\ref{eq:eom}\cite{TRANCHIDA2018406}.

\subsection{Spin-Lattice model for BCC Fe}
\label{section:model_Fe_Ni}

We study the MEL effects on sound velocity for BCC Fe at zero-temperature. To build the spin-lattice model for this material, we follow the procedure described in Ref.\cite{nieves2021spinlattice_prb}. For the classical interatomic potential $\mathcal{V}(r_{ij})$, we use the spectral neighbor analysis potential (SNAP)~\cite{thompson2015spectral} for BCC Fe developed by Nikolov et al.\cite{Nikolov2021}, that yields very good agreement with first-principles calculations. 
The magneto-elastic contribution to this potential is modified in order to improve its predictions of magnetostrictive properties.
The calculated parameters of the Bethe-Slater curve for $J(r_{ij})$ and the N\'{e}el dipole and quadrupole terms are given in Table \ref{table:data_BS}. We use an effective short range parameterization of the Bethe-Slater curve\cite{nieves2021spinlattice_prb} for $J(r_{ij})$ in order to reproduce experimental Curie temperature $T_C=1043$ K\cite{Handley} and theoretical volume magnetostriction $\omega_s=(V_0^c-V_0^r)/V_0^r$ (where $V_0^c$ and $V_0^r$ are the equilibrium volumes at the collinear state and paramagnetic-like state, respectively) calculated by Shimizu using the itinerant electron model\cite{shimizu1978}, that is $\omega_s=1.16\times10^{-2}$. From the analysis of the equation of state (EOS) we obtain that the equilibrium lattice parameter for the collinear state without SOC of BCC Fe is $a_0^c=2.83023$ \r{A}, and for the state with randomly oriented magnetic moments (paramagnetic-like state) is $a_0^r=2.81937$ \r{A}. The elastic constants ($C_{ij}$), anisotropic magnetostrictive coefficients and MCA given by this model are shown in Table \ref{table:data_properties}, which are calculated combining the AELAS\cite{AELAS} and MAELAS\cite{maelas_publication2021} packages, both interfaced with LAMMPS\cite{nieves2021spinlattice_prb}. The elastic constants for the state with randomly oriented magnetic moments are calculated using a supercell size of $30\times30\times30$ unit cells ($54000$ atoms). For the magnetic moment we use the experimental values at zero pressure and zero temperature $\mu_{Fe}=2.2\mu_B$ \cite{Handley}, while for the mass we use $55.85$ g/mole.

\begin{table}[h!]
\caption{Parameters of the SD-MD model for BCC Fe.}
\label{table:data_BS}
\centering
\begin{tabular}{cc}
\toprule
\begin{tabular}[c]{@{}c@{}}\textbf{SD-MD model} \\ \textbf{parameters}\end{tabular}		& \quad\textbf{BCC Fe} \quad\quad \\
\midrule
$\alpha_l$ ($\mu$eV/atom)					& 377.32   \\
$\gamma_l$                      & 0.78979  \\
$\delta_l$ (\r{A})					& 2.45105   \\
$R_{c,l}$ (\r{A})					& 2.6  \\
\midrule
$\alpha_q$	($\mu$eV/atom)					& 29.965  \\
$\gamma_q$                      & 1.0496  \\
$\delta_q$	(\r{A})				& 2.45105  \\
$R_{c,q}$ (\r{A})					& 2.6  \\
\midrule
$\alpha_J$ (meV/atom)					& -14.2048  \\
$\gamma_J$                      & 2.6125   \\
$\delta_J$ (\r{A})					&  2.45105  \\
$R_{c,J}$ (\r{A})					& 2.6  \\
\bottomrule
\end{tabular}
\end{table}

\begin{table*}[ht]
\caption{Calculated and experimental elastic constants, MEL constants ($b_i$), magnetostrictive coefficients ($\lambda$),  MCA ($K_1$), saturation magnetization ($M_s$) and density ($\rho$) for BCC Fe at zero-temperature. The elastic constants are calculated assuming collinear ($C^c_{ij}$) and randomly ($C^r_{ij}$) oriented atomic magnetic moments without SOC, as explained in the main text.}
\label{table:data_properties}
\centering
\resizebox{\textwidth}{!}{
\begin{tabular}{cccc|ccc|ccc|cc|cc|cc}
\toprule
 \begin{tabular}[c]{@{}c@{}}\textbf{Elastic} \\ \textbf{constants}\end{tabular}	& \begin{tabular}[c]{@{}c@{}}$C^c_{ij}$\\\textbf{SD-MD} \\ \textbf{(GPa)}\end{tabular}	
& \begin{tabular}[c]{@{}c@{}}$C^r_{ij}$\\\textbf{SD-MD} \\ \textbf{(GPa)}\end{tabular}
& \begin{tabular}[c]{@{}c@{}}\textbf{Expt.} \\ \textbf{(GPa)}\end{tabular}  &
$b$	& \begin{tabular}[c]{@{}c@{}}\textbf{SD-MD} \\ \textbf{(MPa)}\end{tabular}	& \begin{tabular}[c]{@{}c@{}}\textbf{Expt.} \\ \textbf{(MPa)}\end{tabular}  &
$\lambda$	& \begin{tabular}[c]{@{}c@{}}\textbf{SD-MD} \\ \textbf{($\times10^{-6}$)}\end{tabular}	& \begin{tabular}[c]{@{}c@{}}\textbf{Expt.} \\ \textbf{($\times10^{-6}$)}\end{tabular}  	& \begin{tabular}[c]{@{}c@{}}$K_1$\\\textbf{SD-MD} \\ \textbf{(kJ/m$^3$)}\end{tabular}	& \begin{tabular}[c]{@{}c@{}}$K_1$\\\textbf{Expt.} \\ \textbf{(kJ/m$^3$)}\end{tabular} &  	 \begin{tabular}[c]{@{}c@{}}$\mu_0M_s$\\\textbf{SD-MD} \\ \textbf{(T)}\end{tabular}	& \begin{tabular}[c]{@{}c@{}}$\mu_0M_s$\\\textbf{Expt.} \\ \textbf{(T)}\end{tabular} 
     & \begin{tabular}[c]{@{}c@{}}$\rho^c$ \\\textbf{SD-MD} \\ \textbf{(g/cm$^3$)} \end{tabular} &  \begin{tabular}[c]{@{}c@{}}$\rho^r$ \\\textbf{SD-MD} \\ \textbf{(g/cm$^3$)} \end{tabular} \\

\midrule
 $C_{11}$ & 252.56 & 256.60 & 243$^a$  & $b_1$ & -4.41 & -3.3$^e$ &  $\lambda_{001}$ & 26.08 & 26$^c$ &   55.001  &  55$^d$ & 2.26 & 2.19$^c$ & 8.18 &	8.28		\\
                   $C_{12}$ & 139.87 & 143.84 & 138$^a$ & $b_2$ & 9.73 & 10.5$^e$ &  $\lambda_{111}$ & -30.33 & -30$^c$ &   &  &  &  \\
			$C_{44}$ & 106.95 & 119.99 & 122$^a$ & & & &  &  & & & & & \\

\bottomrule
$^a$Ref.\cite{Fe_elas_exp}, & $^b$Ref.\cite{Lee2003}, & $^c$Ref.\cite{Handley}, \\ $^d$Ref.\cite{Getz} & $^e$Ref.\cite{Burkert}
\end{tabular}
}
\end{table*}

\subsection{Computational calculation of sound velocity}
\label{section:sound_velocity}

To calculate the sound velocity, we use three different approaches based on: i) the oscillation of the kinetic energy\cite{Cai}, (ii) the finite displacement method\cite{TOGO20151} (FDM) and (iii) corrections to elastic constants\cite{Rinaldi1985}. Below we provide some details about these methods.

\subsubsection{Oscillation of the kinetic energy}
\label{section:Ek}

\begin{figure}[h]
\centering
\includegraphics[width=0.6\columnwidth ,angle=0]{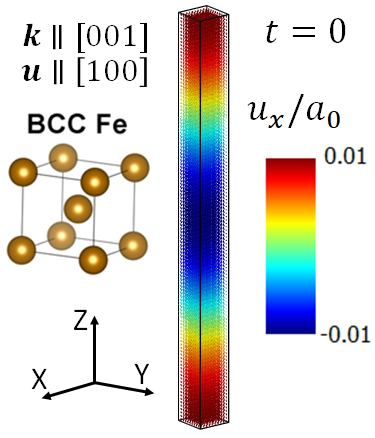}
\caption{Initial atomic displacement of a transverse phonon mode propagating in the direction $\boldsymbol{k}_{ph}\parallel[001]$ with polarization $\boldsymbol{u}\parallel[100]$ for BCC Fe. The system size is $10\times10\times120$ unit cells, while the displacement amplitude is $u_0=0.01a_0$. The phonon momentum is $k_{ph}=2\pi/(120 a_0)$.}
\label{fig:sys}
\end{figure}

\begin{figure}[h]
\centering
\includegraphics[width=\columnwidth ,angle=0]{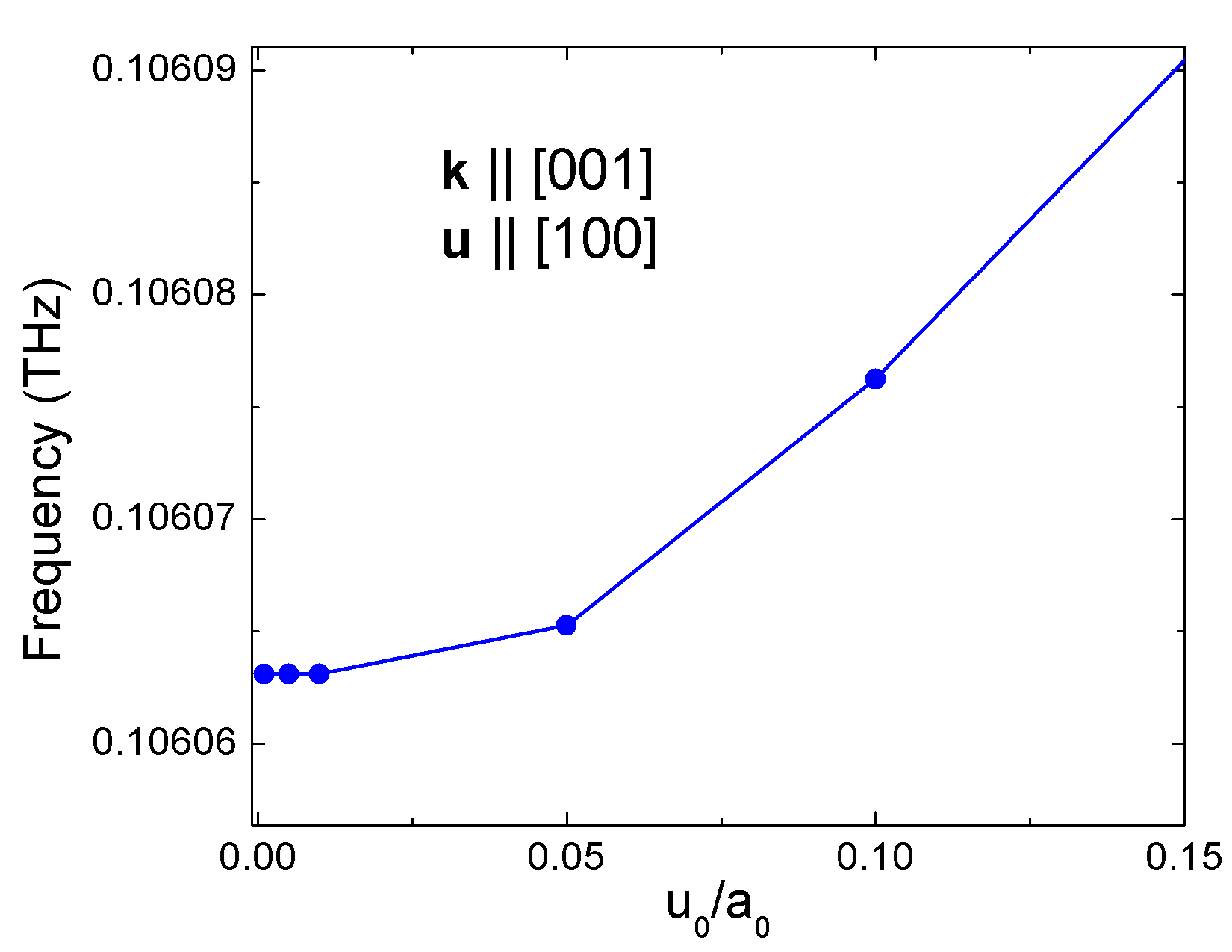}
\caption{Calculated frequency versus the initial displacement amplitude $u_0$ of a phonon propagating in the direction $\boldsymbol{k}_{ph}\parallel[001]$ with polarization $\boldsymbol{u}\parallel[100]$ for BCC Fe at the collinear state without SOC.}
\label{fig:f_vs_u0}
\end{figure}

\begin{figure}[h]
\centering
\includegraphics[width=\columnwidth ,angle=0]{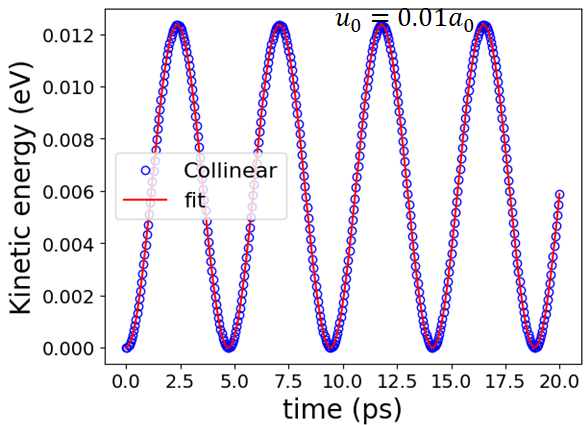}
\caption{ Fitting to extract the frequency from the oscillation of the kinetic energy of a transverse standing wave propagating in the direction $\boldsymbol{k}\parallel[001]$ with polarization $\boldsymbol{u}\parallel[100]$ for BCC Fe at the collinear state without SOC. Blue points stand for the calculation with spin-lattice simualtions, while the red line represents the fitting to Eq.\ref{eq:eK}. }
\label{fig:fit_fo}
\end{figure}

The first method makes use of the oscillation of the kinetic energy to derive the sound velocity\cite{Cai}. Namely, as initial condition for spin-lattice dynamics, one displaces the atoms to generate a standing plane wave with sufficiently large wavelength (isolated phonon with low momentum $\boldsymbol{k}_{ph}$)
\begin{equation}
\begin{aligned}
\boldsymbol{u}(\boldsymbol{r},t)=\boldsymbol{u}_0\cos(\boldsymbol{k}_{ph}\cdot\boldsymbol{r})\cos(2\pi f_{ph} t),
\label{eq:wave}
\end{aligned}
\end{equation}
where $\boldsymbol{u}$ is the displacement vector, $\boldsymbol{u}_0$ is the displacement amplitude, $f_{ph}$ is the frequency of the phonon, $\boldsymbol{r}$ is the position and $t$ is the time. Here, periodic boundary conditions are used, and $k_{ph}=2\pi n/L$, where $n$ is an integer and $L$ is the length of the simulated system along $\boldsymbol{k}_{ph}$. Next, one runs spin-lattice dynamics (with initial velocities of atoms equal to zero to simulate sound velocity at zero-temperature) using the microcanonical ensemble NVE for at least few periods of the kinetic energy. In our simulation we set the time step $dt = 1$ fs, and we verify that the total energy is preserved. Last, the phonon frequency $f_{ph}$ is extracted from the fitting of the kinetic energy ($E_K$) versus time\cite{Cai}
\begin{equation}
\begin{aligned}
E_K(t)=A\left[1-\cos(4\pi f_{ph} t)\right].
\label{eq:eK}
\end{aligned}
\end{equation}
where $A$ is a fitting parameter that does not depend on time since no attenuation of the sound wave takes place due to the lack of energy dissipation for these particular simulation conditions. Here, in the limit of low momentum ($k_{ph}\xrightarrow{}0$), the phonon's velocity (group velocity) $v_{ph}$ approaches the sound velocity $v$ in the solid (continuum theory) that can be expressed in terms of the elastic constants. For example, in the case of a sound wave propagating in the direction $[001]$, one can approximately compute the sound velocity in the solid from the phonon's frequency and momentum  as
\begin{equation}
\begin{aligned}
v\Bigg\vert_{\boldsymbol{u}}^{\boldsymbol{k}\parallel[001]}=\lim_{k_{ph}\rightarrow 0}v_{ph}\Bigg\vert_{\boldsymbol{u}}^{\boldsymbol{k}_{ph}\parallel[001]}=\frac{2\pi f_{ph}}{k_{ph}}\Bigg\vert_{\boldsymbol{u}}^{\boldsymbol{k}_{ph}\parallel[001]}.
\label{eq:v}
\end{aligned}
\end{equation}
 In our calculations, we use a system size $10\times10\times120$ unit cells ($24000$ atoms, $L=120 a_0$) with lattice parameter $a_0^c$ and low phonon momentum $n=1$ ($k_{ph}=2\pi/(120 a_0)$) for BCC Fe at collinear state, see Fig.\ref{fig:sys}.  In the case of paramagnetic-like state, we increase the system size up to $30\times30\times120$ unit cells ($216000$ atoms) with lattice parameter $a_0^r$, as discussed in Section \ref{subsection:iso}. The simulations are performed using the SPIN package\cite{TRANCHIDA2018406} of LAMMPS\cite{thompson2022lammps}. As first benchmark  of this method, we study the influence of the displacement amplitude $\boldsymbol{u}_0$ on the calculated frequency for a  transverse phonon propagating in the direction $\boldsymbol{k}_{ph}\parallel[001]$ with polarization $\boldsymbol{u}\parallel[100]$. The results are  shown in Fig.\ref{fig:f_vs_u0}. We see that below $u_0=0.01a_0$ the frequency is not significantly affected by this parameter, so we use this value for our calculations. The fitting to extract the frequency for this case (collinear state without SOC) is shown in Fig.\ref{fig:fit_fo}. The calculated velocity using Eq.\ref{eq:v} with the fitted frequency $f_{ph}^c$ and the value of $k_{ph}^c=2\pi/(120 a_0^c)$ is $v_{ph}^c=3602.2$ m/s. Note that in this method to further reduce the phonon's momentum ($k_{ph}\xrightarrow{}0$) one needs to increase the system size ($L$) making the simulations more demanding computationally. On the other hand, one advantage of this method is that could be used to study the MEL effects on the attenuation of sound waves\cite{Cai,Simon+1958+84+89,sakurai,booktremolet} through the time dependence of the fitting parameter $A(t)$ in Eq.\ref{eq:eK} within a simulation that allows energy dissipation. The study of MEL effects on the attenuation of sound waves is not performed in the present work.

 \subsubsection{Finite displacement method}
\label{section:fdm}
 
To overcome the size limitations of the method based on the oscillation of the kinetic energy, one can use the FDM\cite{TOGO20151}. In the FDM, the phonon's frequency is derived from the forces associated
with a systematic set of displacements. Here, we use this method, as implemented in the program PHONOPY\cite{TOGO20151,phonopy}, also leveraging the PHONOLAMMPS\cite{phonolammps} interface between PHONOPY and LAMMPS.  The group velocity of the phonon is computed as\cite{phonopy}
\begin{equation}
\begin{aligned}
v_{ph}(\boldsymbol{k},j)=\frac{1}{4\pi f_{ph}(\boldsymbol{k},j)}\Big\langle\boldsymbol{e}(\boldsymbol{k},j)\Big\vert\frac{\partial D(\boldsymbol{k})}{\partial \boldsymbol{k}} \Big\vert\boldsymbol{e}(\boldsymbol{k},j)\Big\rangle,
\label{eq:v_fdm}
\end{aligned}
\end{equation}
where $j$ is the phonon's band index, $D$ is the dynamical matrix and $\boldsymbol{e}$ is the phonon  polarization vector. For the collinear calculations we generate a supercell with size $4\times4\times4$ with lattice parameter $a_0^c$, while for the paramagnetic states we use $16\times16\times16$ with lattice parameter $a_0^r$. The atomic displacement distance is set to its default value $0.01$ \r{A}. Since in this method we do not have size limitations for studying phonons with very low momentum ($k_{ph}\xrightarrow{}0$), we compute the frequency and group velocity for a phonon with momentum $k_{ph}=2\pi/(12000 a_0)$. For example, in the case of a transverse phonon at a collinear state propagating in the direction $\boldsymbol{k}_{ph}\parallel[001]$ with very low momentum $\boldsymbol{k}=(0,0,2\pi/(12000 a_0^c))$, the calculated group velocity with the FDM is $v_{ph}^c=3601.2$ m/s, which is quite close to the value obtained with the oscillation of the kinetic energy ($v_{ph}^c=3602.2$ m/s).

 \subsubsection{Corrections to elastic constants}
\label{section:correc_elas}
 
 As third method, we consider the MEL effects as corrections to elastic constants. A detailed description of this approach is given in Appendix \ref{app:eff_cij}. For example,  in the case of a  transverse sound wave propagating in the direction $\boldsymbol{k}_{ph}\parallel[001]$ with polarization $\boldsymbol{u}\parallel[100]$ at a collinear state without SOC, we have
\begin{equation}
\begin{aligned}
v^c\Bigg\vert_{\boldsymbol{u}\parallel[100]}^{\boldsymbol{k}\parallel[001]}=\sqrt{\frac{C^c_{44}}{ \rho^c}},
\label{eq:v0}
\end{aligned}
\end{equation}
where $\rho^c$ is the density of the material at the collinear state. Using the elastic constant $C^c_{44}$ and density $\rho^c$ (see Table \ref{table:data_properties}) at the collinear state in Eq.\ref{eq:v0}, we find $v^c=3615.7$ m/s. Although slightly larger than the previously simulated phonon's velocity with the oscillation of the kinetic energy and FDM, this remains within a close agreement.

\section{Theory of magnetoelastic effects on sound velocity}
\label{section:theory}

In this section we provide a brief overview of the main theoretical isotropic and anisotropic MEL effects on sound velocity. To facilitate the analysis of the simulations, it is convenient to decompose the isotropic and anisotropic MEL effects on the fractional change in velocity as follows
\begin{equation}
\begin{aligned}
& \frac{v^{SOC}-v^{r}}{v^{r}}  =\frac{v^{c}-v^{r}}{v^{r}}+\frac{v^{SOC}-v^{c}}{v^{r}} \\
& =\frac{v^{c}-v^{r}}{v^{r}} + \frac{v^{SOC}-v^{c}}{v^{c}}\cdot \frac{1}{1-\left(\frac{v^c-v^r}{v^{c}}\right)}\\
& = \frac{v^{c}-v^{r}}{v^{r}} + \frac{v^{SOC}-v^{c}}{v^{c}}\cdot \left[1+\left(\frac{v^c-v^r}{v^{c}}\right)+...\right]\\
& \simeq \frac{v^{c}-v^{r}}{v^{r}} + \frac{v^{SOC}-v^{c}}{v^{c}}+O\left[\left(\frac{v^{SOC}-v^r}{v^{c}}\right)\cdot\left(\frac{v^c-v^r}{v^{c}}\right)\right],
\label{eq:df}
\end{aligned}
\end{equation}
where $v^{SOC}$ is the sound velocity at a collinear state including  isotropic exchange and SOC interactions, $v^{r}$ is the sound velocity assuming randomly oriented atomic magnetic moments (paramagnetic-like state) including  only isotropic exchange interaction, and $v^{c}$ is the sound velocity at a collinear state including  only isotropic exchange interaction. In the last step, we assumed that the change in the sound velocity due to MEL effects is small in comparison to the sound velocity ($\vert v^{SOC}-v^c\vert/v^c\ll 1$ , $\vert v^c-v^r\vert /v^r\ll 1$), which is a reasonable approximation for the material studied here. 

In experiment, the sound velocity is typically measured through the pulse echo method\cite{TRUELL196953}. The measured frequency with this technique $F$ is related to the sound velocity $v$ and sample length $l$ along the direction of the wave propagation through the following equation\cite{TRUELL196953,Thurston,ROUCHY198159}
\begin{equation}
\begin{aligned}
v(F,l)=2Fl.
\label{eq:d_f}
\end{aligned}
\end{equation}
Performing a Taylor expansion of sound velocity yields
\begin{equation}
\begin{aligned}
v(F,l) & =v(F_0,l_0)+\left(\frac{\partial v}{\partial F}\right)_{l=l_0}(F-F_0)+\left(\frac{\partial v}{\partial l}\right)_{F=F_0}(l-l_0)\\
& = v_0 +2l_0(F-F_0)+2F_0(l-l_0),
\label{eq:dv_f}
\end{aligned}
\end{equation}
where $v(F_0,l_0)=v_0=2F_0l_0$. If we subtract $v_0$ from both sides of Eq.\ref{eq:dv_f} and divide both sides by $v_0$, then we obtain\cite{Rouchy1979,booktremolet}
\begin{equation}
\begin{aligned}
\frac{v-v_0}{v_0}=\frac{F-F_0}{F_0}+\frac{l-l_0}{l_0}.
\label{eq:dv_f_2}
\end{aligned}
\end{equation}
The theoretical expressions for the fractional change in pulse echo frequency $F$ have been derived by Rouchy et al.\cite{Rouchy1979,ROUCHY198069,booktremolet}. In addition to this contribution, in Eq.\ref{eq:dv_f_2} we see that the fractional change in length along the direction of wave propagation must be also calculated in order to compute the fractional change in velocity. This fractional change in length arises from the magnetostriction induced by MEL interactions, and its general form for an arbitrary measuring length direction $\boldsymbol{\beta}$ is given by Eq.\ref{eq:delta_l_cub_I}.

The first term in the right-hand side of Eq.\ref{eq:df} corresponds to the fractional change in sound velocity due to the isotropic exchange interaction (isotropic MEL effects)\cite{booktremolet,Rouchy1979}
\begin{equation}
\begin{aligned}
\frac{v^{c}-v^{r}}{v^{r}}\Bigg\vert_{\boldsymbol{u}}^{\boldsymbol{k}} & = \frac{F^c-F^r}{F^r}\Bigg\vert_{\boldsymbol{u}}^{\boldsymbol{k}}+\frac{l^c-l^r}{l^r}\Bigg\vert^{\boldsymbol{k}},\\
\frac{F^c-F^r}{F^r}\Bigg\vert_{\boldsymbol{u}}^{\boldsymbol{k}} & = Y(m^{\alpha,2}_i)\Bigg\vert_{\boldsymbol{u}}^{\boldsymbol{k}},
\label{eq:dfiso}
\end{aligned}
\end{equation}
where $Y(m^{\alpha,2}_i)$ is a function that depends on the isotropic morphic coefficients $m^{\alpha,2}_i$. The form of this function also depends on the propagation direction of the wave $\boldsymbol{k}$ and polarization $\boldsymbol{u}$ but not on the magnetization direction $\boldsymbol{M}$, since this MEL effect is isotropic. In this isotropic case, the fractional change in length depends only  on the propagation direction of the wave. The second term in the right hand side of  Eq.\ref{eq:df} gives the anisotropic MEL effects due to the SOC, and may be splitted into the following terms\cite{booktremolet,Rouchy1979,ROUCHY198069}
\begin{equation}
\begin{aligned}
 \frac{v^{SOC}-v^{c}}{v^{c}}\Bigg\vert_{\boldsymbol{u}}^{\boldsymbol{k},\boldsymbol{M}} & = \frac{F^{SOC}-F^c}{F^c}\Bigg\vert_{\boldsymbol{u}}^{\boldsymbol{k},\boldsymbol{M}}+\frac{l^{SOC}-l^c}{l^c}\Bigg\vert^{\boldsymbol{k},\boldsymbol{M}},\\
\frac{F^{SOC}-F^c}{F^c}\Bigg\vert_{\boldsymbol{u}}^{\boldsymbol{k},\boldsymbol{M}} & = G(m^{\gamma,2}_i,m^{\epsilon,2}_i)\Bigg\vert_{\boldsymbol{u}}^{\boldsymbol{k},\boldsymbol{M}}+ R(\lambda^{\gamma,2},\lambda^{\epsilon,2})\Bigg\vert_{\boldsymbol{u}}^{\boldsymbol{k},\boldsymbol{M}}\\
 & +S(H)\Bigg\vert_{\boldsymbol{u}}^{\boldsymbol{k},\boldsymbol{M}},
\label{eq:dfani}
\end{aligned}
\end{equation}
where $G(m^{\gamma,2}_i,m^{\epsilon,2}_i)$ is a function that depends on anisotropic morphic coefficients $m^{\gamma,2}_i$ and $m^{\epsilon,2}_i$ which are linear combinations of second order in strain MEL constants and third order in strain elastic constants, $R(\lambda^{\gamma,2},\lambda^{\epsilon,2})$ is a function that comes from  rotational and magnetostrictive effects\cite{ROUCHY198069} and depends on anisotropic magnetostrictive coefficients $\lambda^{\gamma,2}$ and $\lambda^{\epsilon,2}$, and $S(H)$ is a field-dependent term derived by Simon \cite{Simon+1958+84+89}. The form of these functions also depends on $\boldsymbol{k}$, $\boldsymbol{u}$ and $\boldsymbol{M}$. The fractional change in length depends on $\boldsymbol{k}$ and $\boldsymbol{M}$ but not on the polarization $\boldsymbol{u}$. The relationship between the morphic coefficients and higher order elastic and MEL constants are given in Eq.\ref{eq:morphic_coeff}, where we use the same definitions as in Ref.\cite{ROUCHY198069}. Eqs.\ref{eq:dfiso} and \ref{eq:dfani} are derived from the free energy by solving the
coupled elastic and magnetic equations of motion \cite{Simon+1958+84+89,Rouchy1979,ROUCHY198069,Ikeda_1982,booktremolet}. A theoretical 
description of all these effects can also be provided by expanding the internal energy as a series of
the Lagrangian tensor components, as well as symmetrical and
antisymmetrical components of the homogeneous
strains \cite{Rouchy1979}. The explicit form of these equations for the simulated cases is shown in Section \ref{section:results}. Alternatively, one can also take into account MEL effects on sound velocity as corrections to the elastic constants \cite{Rinaldi1985,booktremolet}. We describe this approach in Appendix \ref{app:eff_cij}.

\section{Results}
\label{section:results}

\subsection{Isotropic magnetoelastic effects}
\label{subsection:iso}

\begin{figure}[h]
\centering
\includegraphics[width=\columnwidth ,angle=0]{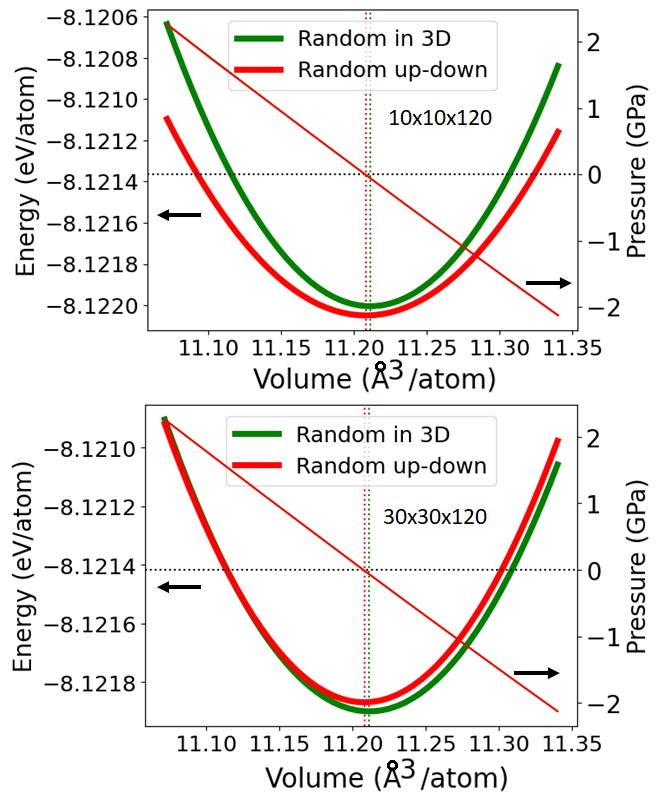}
\caption{Calculated equation of state of BCC Fe assuming two types of paramagnetic-like state: (green line) with randomly oriented magnetic moments along 3D directions, and (red line) random up-down orientations. Vertical dash line stands for the equilibrium volume $V_0^r$. The system size is (top) $10\times10\times120$ and (bottom) $30\times30\times120$ unit cells.}
\label{fig:eos_para}
\end{figure}

\begin{figure}[h]
\centering
\includegraphics[width=\columnwidth ,angle=0]{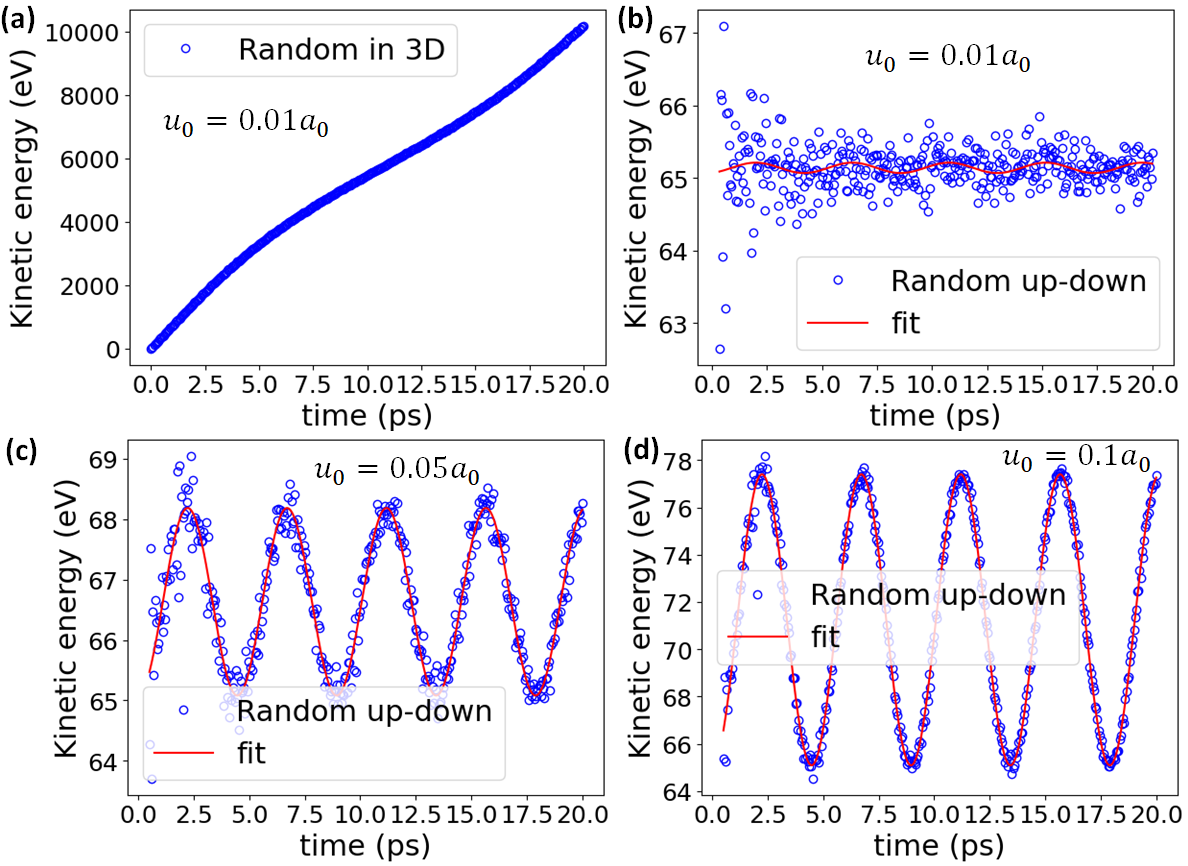}
\caption{ Kinetic energy versus time of a transverse standing wave propagating in the direction $\boldsymbol{k}\parallel[001]$ with polarization $\boldsymbol{u}\parallel[100]$ for BCC Fe at paramagnetic-like states: (a) with randomly oriented magnetic moments along 3D directions and random up-down orientations using an initial displacement amplitude (b) $u_0=0.01a_0$, (c) $u_0=0.05a_0$ and (d) $u_0=0.1a_0$.}
\label{fig:for_Fe}
\end{figure}

In the case of a transverse wave propagating along $\boldsymbol{k}\parallel[001]$ with polarization $\boldsymbol{u}\parallel[100]$,  Eq.\ref{eq:dfiso}  reads\cite{Rouchy1979}
\begin{equation}
\begin{aligned}
\frac{v^{c}-v^{r}}{v^{r}}\Bigg\vert_{\boldsymbol{u}\parallel[100]}^{\boldsymbol{k}\parallel[001]} & = \frac{F^{c}-F^{r}}{F^{r}}\Bigg\vert_{\boldsymbol{u}\parallel[100]}^{\boldsymbol{k}\parallel[001]} +\frac{l^{c}-l^{r}}{l^{r}}\Bigg\vert^{\boldsymbol{k}\parallel[001]},\\
\frac{F^{c}-F^{r}}{F^{r}}\Bigg\vert_{\boldsymbol{u}\parallel[100]}^{\boldsymbol{k}\parallel[001]} & = Y(m_i^{\alpha,2})\Bigg\vert_{\boldsymbol{u}\parallel[100]}^{\boldsymbol{k}\parallel[001]}  = \frac{m_3^{\alpha,2}}{12C^c_{44}},\\
\frac{l^{c}-l^{r}}{l^{r}}\Bigg\vert^{\boldsymbol{k}\parallel[001]} & = \frac{a_0^c-a_0^r}{a_0^r},
\label{eq:dfk001u100m001iso}
\end{aligned}
\end{equation}
where $m_3^{\alpha,2}$ is an isotropic morphic coefficient defined in Eq.\ref{eq:morphic_coeff}. The form of function $Y$ is the same as in Ref.\cite{Rouchy1979}, but note that it is not the same as in Ref.\cite{booktremolet} due to a different definition of morphic coeffcients and high order MEL constants. The fractional change in length along $[001]$ arises from volume magnetostriction $\omega_s$ due to the exchange interaction.  Here, to calculate $v^r$ we consider two types of paramagnetic-like state: (i) with randomly oriented magnetic moments along 3D directions, and (ii) random up-down orientations\cite{Alling}. To simulate the purely isotropic MEL effects, we do not include SOC interaction (dipole and quadrupole N\'{e}el's terms) in the spin-lattice model. In Fig.\ref{fig:eos_para} we show the calculated EOS of BCC Fe at these paramagnetic-like states using  system sizes $10\times10\times120$ and $30\times30\times120$ unit cells. We observe that increasing the system size up to $30\times30\times120$ gives similar EOS for both types of paramagnetic-like states, so that we use this size to calculate $v^r$ through the method based on the oscillation of the kinetic energy. Moreover, since this system size is sufficiently large, we do not need to compute $v^r$ for several random configurations and average them, or use special quasi-random structure (SQS) method\cite{Alling}. Note that for the collinear state we do not need such a large size, so that we use a system size $10\times10\times120$ unit cells to compute $v^c$. In Fig.\ref{fig:for_Fe} we show the calculated kinetic energy versus time for BCC Fe at these two considered paramagnetic-like states. Here, we simulated three cases using an initial displacement amplitude $u_0=0.01a_0$, $u_0=0.05a_0$ and $u_0=0.1a_0$. We see that the paramagnetic-like state with randomly oriented magnetic moments along 3D directions increases the kinetic energy rapidly, so that it is not possible to fit this behaviour to Eq. \ref{eq:eK}. This is due to the fact that this spin configuration corresponds to a thermal equilibrium state with spin temperature $T_{spin}$ above $T_C=1043 K$ ($T_{spin}>T_C$), hence the spin system interchanges energy with the phonon system, which is initially at zero-temperature ($T_{ph}=0$ K), trying to equilibrate their  temperatures\cite{TRANCHIDA2018406}. Consequently, the lattice temperature increases rapidly, as seen in the dynamics of the kinetic energy. On the other hand, the configuration with random up-down spins is a paramagnetic-like state in terms of the total energy but it does not correspond to a thermally equilibrated state above $T_C$. For example, the spin temperature for this state using the definition of Nurdin et al.\cite{Nurdin,TRANCHIDA2018406} is $T_{spin}=0$ K. Hence, it can mechanically stabilize the oscillation of the kinetic energy by introducing a small energy shift ($\sim 65$ eV $=0.3$ meV/atom) on the kinetic energy (Eq.\ref{eq:eK}). In this case, we can extract the frequency of the sound wave $f^r$ (do not confuse with the measured frequency $F$ in the pulse echo method) if the initial displacement amplitude $u_0$ is sufficiently large ($u_0>0.01a_0$) to overcome the noise fluctuations of the kinetic energy. From the fitting we obtain the frequency $f^r=0.111619$ THz. We can estimate the sound velocity at the paramagnetic state by inserting the calculated $f^r$ and $k^r=2\pi/(120a_0^r)$ in Eq.\ref{eq:v}, finding $v^r=3776.4$ m/s. At the collinear state, we obtain the frequency $f^c=0.106063$ THz. Hence,  the estimated sound velocity using Eq.\ref{eq:v} is  $v^c=3602.2$ m/s. Consequently, using Eq.\ref{eq:v} leads to a fractional change in velocity $(v^c-v^r)/v^r=-0.0435$.

The FDM is a more rigorous and accurate approach to compute $v^r$ than the oscillation of the kinetic energy since it has no problems associated with the spin-phonon stability for the  paramagnetic-like states at zero-temperature.  Applying the methodology described in Section \ref{section:fdm}, the FDM gives $(v^c-v^r)/v^r=-0.048$ and $-0.0408$ for  paramagnetic-like state with randomly oriented magnetic moments in 3D and randomly up-down orientations, respectively. Alternatively, we can also estimate it by using the corrections to elastic constants, as explained in Section \ref{section:correc_elas} and Appendix \ref{app:eff_cij}. For example, using the elastic constant $C^r_{44}$ and density $\rho^r$ at the paramagnetic state (see Table \ref{table:data_properties}) in Eq.\ref{eq:v0} gives  $v^r= 3807.9$ m/s, while for the collinear state we get  $v^c=3615.7$ m/s. This gives $(v^c-v^r)/v^r=-0.0505$. Therefore, we see that the fractional change in velocity obtained with the three  approaches (from the oscillation of the kinetic energy, FDM and corrections to elastic constants) are consistent with each other. Last, we can also estimate the isotropic morphic coefficient $m_3^{\alpha,2}$ by inserting the calculated sound velocities ($v^r$ and $v^c$), equilibrium lattice parameters ($a^r_0$ and $a^c_0$) and elastic constant $C^c_{44}$ in Eq.\ref{eq:dfk001u100m001iso}. This procedure gives $m_3^{\alpha,2}=-60.8$ GPa using the velocities derived from the oscillation of the kinetic energy, and $m_3^{\alpha,2}=-66.5$ GPa and $-57.4$ GPa with the FDM for  paramagnetic-like state with randomly oriented magnetic moments in 3D and randomly up-down orientations, respectively, while using the velocities given by the corrections to the elastic constants we find $m_3^{\alpha,2}=-69.7$ GPa.

In the case of a longitudinal wave propagating along $\boldsymbol{k}\parallel[001]$ with polarization $\boldsymbol{u}\parallel[001]$,  Eq.\ref{eq:dfiso}  reads\cite{Rouchy1979}
\begin{equation}
\begin{aligned}
\frac{v^{c}-v^{r}}{v^{r}}\Bigg\vert_{\boldsymbol{u}\parallel[001]}^{\boldsymbol{k}\parallel[001]} & = \frac{F^{c}-F^{r}}{F^{r}}\Bigg\vert_{\boldsymbol{u}\parallel[001]}^{\boldsymbol{k}\parallel[001]} +\frac{l^{c}-l^{r}}{l^{r}}\Bigg\vert^{\boldsymbol{k}\parallel[001]},\\
\frac{F^{c}-F^{r}}{F^{r}}\Bigg\vert_{\boldsymbol{u}\parallel[001]}^{\boldsymbol{k}\parallel[001]} & = Y(m_i^{\alpha,2})\Bigg\vert_{\boldsymbol{u}\parallel[001]}^{\boldsymbol{k}\parallel[001]}  = \frac{m_1^{\alpha,2}}{3C^c_{11}},\\
\frac{l^{c}-l^{r}}{l^{r}}\Bigg\vert^{\boldsymbol{k}\parallel[001]} & = \frac{a_0^c-a_0^r}{a_0^r},
\label{eq:dfk001u001m001iso}
\end{aligned}
\end{equation}
where $m_1^{\alpha,2}$ is an isotropic morphic coefficient defined in Eq.\ref{eq:morphic_coeff}. Following the same procedure as before and using Eq.\ref{eq:dfk001u001m001iso}, we find $m_1^{\alpha,2}=-10.6$ GPa and $-8.0$ GPa through the FDM for  paramagnetic-like state with randomly oriented magnetic moments in 3D and randomly up-down orientations, respectively, and $m_1^{\alpha,2}=-4.6$ GPa via the corrections to the elastic constants. Unfortunately, for the simulated cases of the longitudinal mode, the oscillation of the kinetic energy was not sufficiently stable to extract the frequency at the paramagnetic-like state ($f^r$), so that we could not estimate $m_1^{\alpha,2}$ with this approach. In Table \ref{table:data_iso_morphic}, we present a summary of the results obtained for the two isotropic morphic coefficients of BCC Fe calculated in this work. We point out that the elastic constants at the paramagnetic state ($C^r_{ij}$) and the isotropic morphic coefficients ($m^\alpha$) are sensitive to the volume magnetostriction $\omega_S$ of the spin-lattice model.

\begin{table}[h!]
\caption{Calculated isotropic morphic coefficients of a wave propagating along $\boldsymbol{k}\parallel[001]$ using the spin-lattice model of BCC Fe without SOC.}
\label{table:data_iso_morphic}
\centering
\begin{tabular}{ccccc}
\toprule
\begin{tabular}[c]{@{}c@{}}\textbf{Method} \\ \textbf{}\end{tabular} & \begin{tabular}[c]{@{}c@{}}$\boldsymbol{u}$ \\ \textbf{}\end{tabular} & \begin{tabular}[c]{@{}c@{}}\textbf{Paramagnetic} \\ \textbf{State}\end{tabular} &   
\begin{tabular}[c]{@{}c@{}}\textbf{Morphic} \\ \textbf{ coefficient}\end{tabular}		& \quad\textbf{GPa} \quad\\
\midrule
\textrm{Kinetic energy} & 	$\boldsymbol{u}\parallel[100]$		& Random in 3D  &     $m_3^{\alpha,2}$ & - \\
 & 				& Random up-down  &   & -60.8 \\
\textrm{FDM}  &                      & Random in 3D  &    & -66.5 \\
  &                      & Random up-down &     &  -57.4 \\
\textrm{Elastic constants}
& 					& Random in 3D &     & -69.7 \\
\midrule
\textrm{Kinetic energy} & 	$\boldsymbol{u}\parallel[001]$		& Random in 3D  &    $m_1^{\alpha,2}$ & - \\
 & 				& Random up-down  &   & - \\
\textrm{FDM}  &                      & Random in 3D  &   & -10.6\\
  &                      & Random up-down &     & -8.0 \\
\textrm{Elastic constants}
& 					& Random in 3D &    & -4.6\\
\bottomrule
\end{tabular}

\end{table}

\subsection{Anisotropic magnetoelastic effects}
\label{subsection:ani}

\begin{figure}[h]
\centering
\includegraphics[width=\columnwidth ,angle=0]{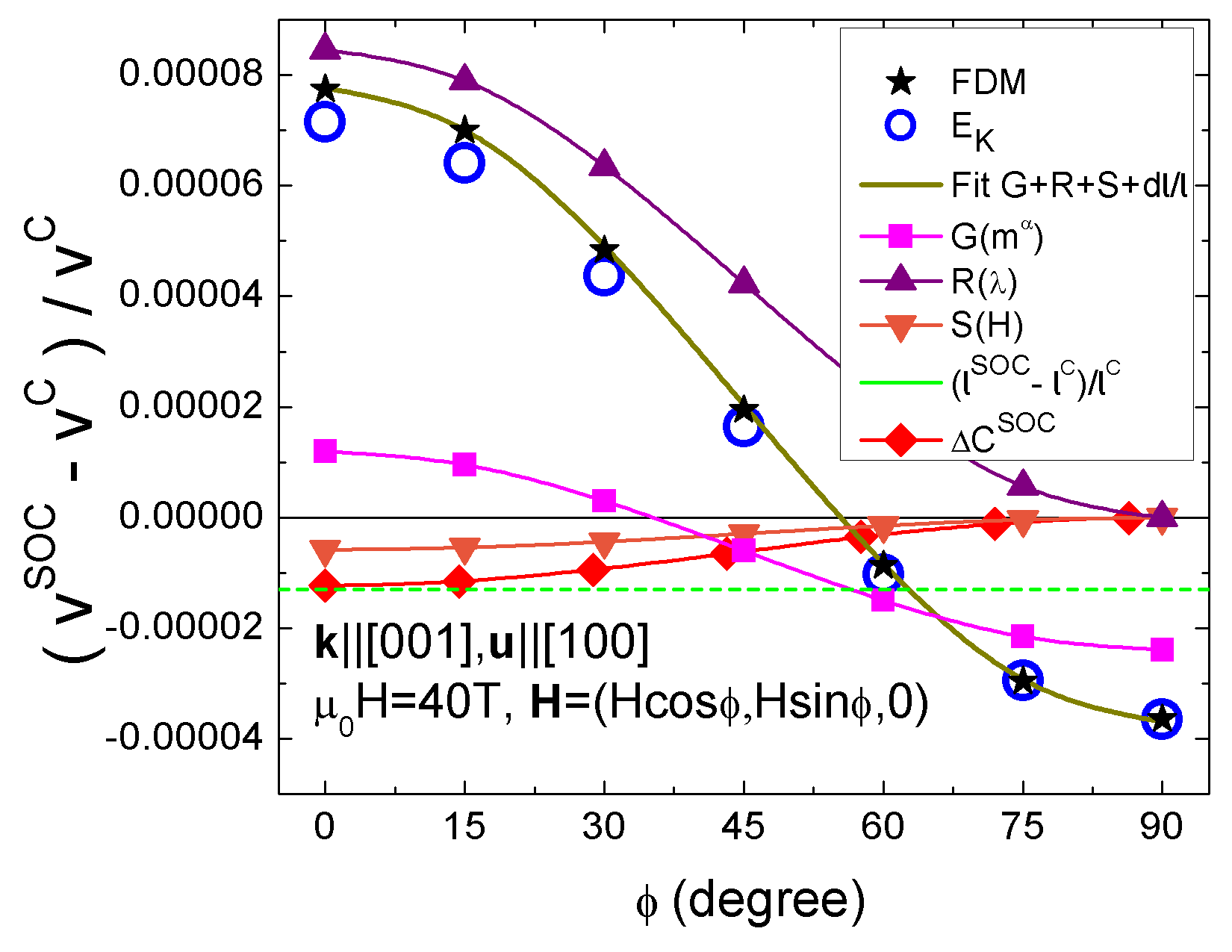}
\caption{Fractional change in velocity of BCC Fe versus the direction of the applied magnetic field $\boldsymbol{H}=(H\cos\phi,H\sin\phi,0)$ on the plane XY for transverse wave propagating along $\boldsymbol{k}\parallel[001]$ with polarization $\boldsymbol{u}\parallel[100]$. The magnitude of the applied field is $\mu_0H=40$ T. Black stars represent the calculations with spin-lattice simulations obtained from the FDM, while brown solid line is the fitting of these data to the summation of functions $G$, $R$, $S$ and $(l^{SOC}-l^c)/l^c$ in Eq.\ref{eq:dfk001u100m100}. Blue circles stand for the calculations with spin-lattice simulations obtained from the oscillation of the kinetic energy ($E_K$). Each function $G$, $R$, $S$ and $(l^{SOC}-l^c)/l^c$ in Eq.\ref{eq:dfk001u100m100} is plotted with pink squares, purple triangles, orange triangles and green dash line, respectively.  Red diamonds correspond to the calculations using the corrections to the elastic constants given by Eq.\ref{eq:v_k001u100_soc}.}
\label{fig:k001u100mxy}
\end{figure}

For the analysis of anisotropic effects we follow the same procedure, but now we switch on the SOC (dipole and quadrupole N\'{e}el's terms) in order to compute the velocity $v^{SOC}$. We first consider a transverse wave propagating along $\boldsymbol{k}\parallel[001]$ with polarization $\boldsymbol{u}\parallel[100]$ and magnetization  $\boldsymbol{M}\bot[001]$. In this case, the functions in Eq. \ref{eq:dfani}  read\cite{ROUCHY198069}
\begin{equation}
\begin{aligned}
& G\Bigg\vert_{\boldsymbol{u}\parallel[100]}^{\boldsymbol{k}\parallel[001]\bot \boldsymbol{M}}  = \frac{m_3^{\gamma,2}}{8C^c_{44}}\left(\frac{1}{3}-\cos2\phi\right),\\ 
& R\Bigg\vert_{\boldsymbol{u}\parallel[100]}^{\boldsymbol{k}\parallel[001]\bot \boldsymbol{M}}  = \frac{1}{2}(\lambda^{\gamma,2}-\lambda^{\epsilon,2})\left[1+\cos2\phi\right]\\
& = \frac{3}{4}(\lambda_{001}-\lambda_{111})\left[1+\cos2\phi\right],\\
& S\Bigg\vert_{\boldsymbol{u}\parallel[100]}^{\boldsymbol{k}\parallel[001]\bot \boldsymbol{M}}  =- \frac{(B^{\epsilon,2})^2(1+\cos2\phi)}{4C^c_{44}\mu_0M_s\left(H+M_s+H_D+\frac{K_1[3+\cos4\phi]}{2\mu_0M_s}\right)},\\ 
& H  \gg\frac{2K_1}{\mu_0M_s},\\
& \frac{l^{SOC}-l^c}{l^c}\Bigg\vert^{\boldsymbol{k}\parallel[001]\bot \boldsymbol{M}}  = -\frac{1}{3}\lambda^{\gamma,2}=-\frac{1}{2}\lambda_{001},
\label{eq:dfk001u100m100}
\end{aligned}
\end{equation}
where $\phi$ is the angle between $\boldsymbol{M}$ and crystallographic direction $[100]$ and $H_D$ is the demagnetizing field. In our simulations $H_D=0$ since we do not include dipole-dipole interactions. The MEL constant $B^{\epsilon,2}=b_2$ and magnetostrictive coefficients $\lambda^{\gamma,2}=3\lambda_{001}/2$ and $\lambda^{\epsilon,2}=3\lambda_{111}/2$ are defined in Appendix \ref{app:Internal_energy}. The form of function $G$ is the same as in Ref.\cite{Rouchy1979}, but note that it is not the same as in Ref.\cite{booktremolet} due to a different definition of morphic coeffcients and high order MEL constants. The fractional change in length along $[001]$ is obtained from the anisotropic part of Eq.\ref{eq:delta_l_cub_I}. The fractional change in velocity obtained from the oscillation of the kinetic energy and the FDM for the spin-lattice model of BCC Fe is shown in Fig.\ref{fig:k001u100mxy}. It is performed at high applied magnetic field $\mu_0H=40$ T where $\boldsymbol{H}\parallel\boldsymbol{M}$. In analogy to experiment, we extract the morphic coeffcient $m_3^{\gamma,2}$ by fitting the results given by the spin-lattice simulations to the summation of functions $G$, $R$, $S$ and $(l^{SOC}-l^c)/l^c$ in Eq.\ref{eq:dfk001u100m100}. Here, all materials parameters in these functions are constrained to corresponding values of the spin-lattice model (see Table \ref{table:data_properties}), except $m_3^{\gamma,2}$ which is a fitting parameter. This procedure gives $m_3^{\gamma,2}=-14.3$ MPa for method based on the oscillation of the kinetic energy, and $m_3^{\gamma,2}=-15.3$ MPa using the FDM. In Fig.\ref{fig:k001u100mxy} we also calculated the fractional change in velocity using the corrections to the elastic constants given by Eq.\ref{eq:v_k001u100_soc}. We see that this approach gives similar results to the field-dependent function $S(H)$, since the linear MEL theory (which does not include high order MEL terms) was used both by Simon\cite{Simon+1958+84+89} to derive function $S(H)$ and Rinaldi et al.\cite{Rinaldi1985} to obtain the  SOC corrections ($\triangle C^{SOC}$) to the elastic constants\cite{booktremolet}.

\begin{figure}[h]
\centering
\includegraphics[width=\columnwidth ,angle=0]{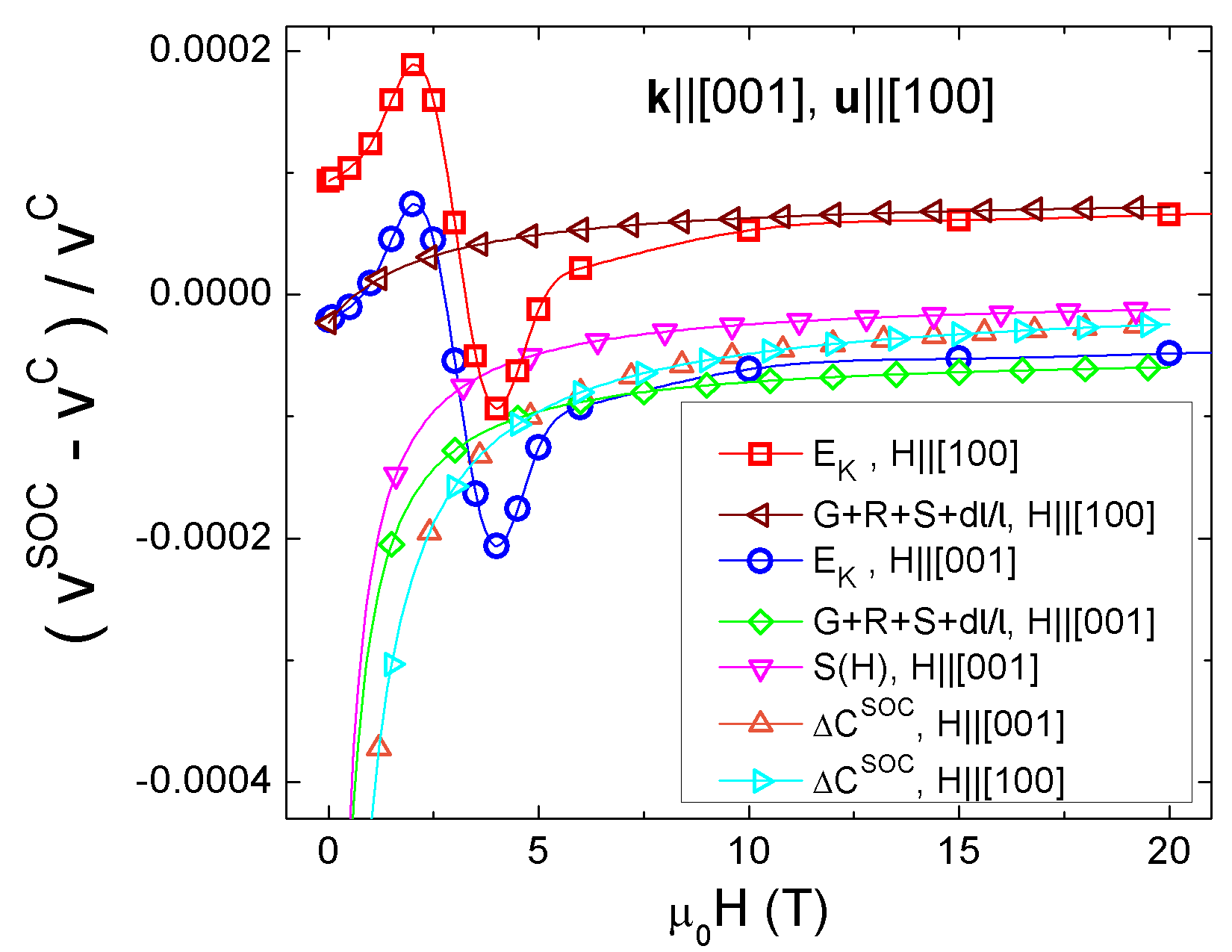}
\caption{Fractional change in velocity of BCC Fe versus the magnitude of the applied magnetic field along $\boldsymbol{H}\parallel[100]$ and $\boldsymbol{H}\parallel[001]$ for a transverse wave propagating along $\boldsymbol{k}\parallel[001]$ with polarization $\boldsymbol{u}\parallel[100]$. Blue points and red squares stand for the calculations with spin-lattice simulations obtained from the oscillation of the kinetic energy for the cases $\boldsymbol{H}\parallel[100]$ and $\boldsymbol{H}\parallel[001]$, respectively. Brown triangles and green diamonds correspond to the summation of functions $G$, $R$, $S$ and $(l^{SOC}-l^c)/l^c$ in Eqs.\ref{eq:dfk001u100m100} and \ref{eq:dfk001u100m001} for the cases $\boldsymbol{H}\parallel[100]$ and $\boldsymbol{H}\parallel[001]$, respectively. The function $S(H)$ in Eq.\ref{eq:dfk001u100m001} is plotted with pink triangles. Orange and cyan triangles represent the calculations using the corrections to the elastic constants given by Eq.\ref{eq:v_k001u100_soc}.}
\label{fig:k001u100m001m100}
\end{figure}

The case of a transverse wave propagating along $\boldsymbol{k}\parallel[001]$ with polarization $\boldsymbol{u}\parallel[100]$ and magnetization  $\boldsymbol{M}\parallel[100]$ is a particular case of Eq.\ref{eq:dfk001u100m001} when $\phi=0$. This case is calculated using the spin-lattice model for different values of the applied field in Fig.\ref{fig:k001u100m001m100}. Here, we also plot the case when the magnetization is $\boldsymbol{M}\parallel[001]$, where the functions in Eq. \ref{eq:dfani}  now read\cite{ROUCHY198069}
\begin{equation}
\begin{aligned}
& G\Bigg\vert_{\boldsymbol{u}\parallel[100]}^{\boldsymbol{k}\parallel\boldsymbol{M}\parallel[001]}  = -\frac{m_3^{\gamma,2}}{12C^c_{44}},\\ & R\Bigg\vert_{\boldsymbol{u}\parallel[100]}^{\boldsymbol{k}\parallel\boldsymbol{M}\parallel[001]}  = \lambda^{\epsilon,2}-\lambda^{\gamma,2}= \frac{3}{2}(\lambda_{111}-\lambda_{001}) ,\\
& S\Bigg\vert_{\boldsymbol{u}\parallel[100]}^{\boldsymbol{k}\parallel\boldsymbol{M}\parallel[001]}  =- \frac{(B^{\epsilon,2})^2}{2C^c_{44}\mu_0M_s\left(H+H_D+\frac{2K_1}{\mu_0M_s}\right)},\quad H\gg\frac{2K_1}{\mu_0M_s},\\
& \frac{l^{SOC}-l^c}{l^c}\Bigg\vert^{\boldsymbol{k}\parallel \boldsymbol{M}\parallel[001]}  = \frac{2}{3}\lambda^{\gamma,2}=\lambda_{001},
\label{eq:dfk001u100m001}
\end{aligned}
\end{equation}
In Fig. \ref{fig:k001u100m001m100}, we see that both calculated cases ($\boldsymbol{M}\parallel[100]$ and $\boldsymbol{M}\parallel[001]$) are in good agreement with the theory at high applied magnetic fields ($\mu_0H>5$ T). In this figure, we used the previously calculated value $m_3^{\gamma,2}=-14.3$ MPa derived from the oscillation of the kinetic energy in order to plot the theoretical function $G$, while for the other parameters in Eqs.\ref{eq:dfk001u100m100} and \ref{eq:dfk001u100m001} we set the values of the spin-lattice model (see Table \ref{table:data_properties}). We also see that the spin-lattice model can correctly reproduce the rotational-magnetostrictive effect at high applied magnetic fields predicted by Eqs.\ref{eq:dfk001u100m100} and \ref{eq:dfk001u100m001}, that is
\begin{equation}
\begin{aligned}
& \lim_{H\to\infty}\left[ \frac{v^{SOC}-v^{c}}{v^{c}}\Bigg\vert_{\boldsymbol{u}\parallel \boldsymbol{M}\parallel[100]}^{\boldsymbol{k}\parallel[001]} -\frac{v^{SOC}-v^{c}}{v^{c}}\Bigg\vert_{\boldsymbol{u}\parallel[100]}^{\boldsymbol{k}\parallel\boldsymbol{M}\parallel[001]}\right]\\
& = \left[R\Bigg\vert_{\boldsymbol{u}\parallel \boldsymbol{M}\parallel[100]}^{\boldsymbol{k}\parallel[001]} +\frac{l^{SOC}-l^c}{l^c}\Bigg\vert^{\boldsymbol{k}\parallel[001]}_{\boldsymbol{M}\parallel[100]}\right]\\
& -\left[R\Bigg\vert_{\boldsymbol{u}\parallel[100]}^{\boldsymbol{k}\parallel\boldsymbol{M}\parallel[001]}+\frac{l^{SOC}-l^c}{l^c}\Bigg\vert^{\boldsymbol{k}\parallel \boldsymbol{M}\parallel[001]} \right] \\
& = 2(\lambda^{\gamma,2}-\lambda^{\epsilon,2}) -\frac{3}{2}\lambda_{001}=\frac{3}{2}\lambda_{001}-3\lambda_{111}.
\label{eq:limit}
\end{aligned}
\end{equation}
The factor $2(\lambda^{\gamma,2}-\lambda^{\epsilon,2})$ arises from the rotational-magnetostrictive effect in the fractional change in pulse echo frequency $F$, as shown in Refs.\cite{Rouchy1979,ROUCHY198069,booktremolet}, while the additional factor $-3\lambda_{001}/2$ comes from the fractional change in length. The fractional change in velocity derived from the corrections to elastic constants (Eq. \ref{eq:v_k001u100_soc}) approaches the field-dependent function $S(H)$ at high fields along $\boldsymbol{H}\parallel[001]$, but it does not reproduce the high order effects since the used corrections to the elastic constants are based on the linear MEL theory\cite{Rinaldi1985}. The results from the spin-lattice simulations exhibit a non-monotonic behaviour at low applied magnetic fields ($\mu_0H<5$ T). Similar pattern has been experimentally observed in single cubic crystals of disordered Co-Pt alloy\cite{ROUCHY198069}. Note that the low-field regime can not be described by the current theory since the field-dependent term $S(H)$ was derived by Simon only under the assumption of high applied magnetic fields\cite{Simon+1958+84+89} ($H\gg2K_1/[\mu_0M_s]$). Hence, we see that the spin-lattice simulations could be a useful tool to explore and understand the physics of MEL effects on sound velocity in this regime.

\begin{figure}[h]
\centering
\includegraphics[width=\columnwidth ,angle=0]{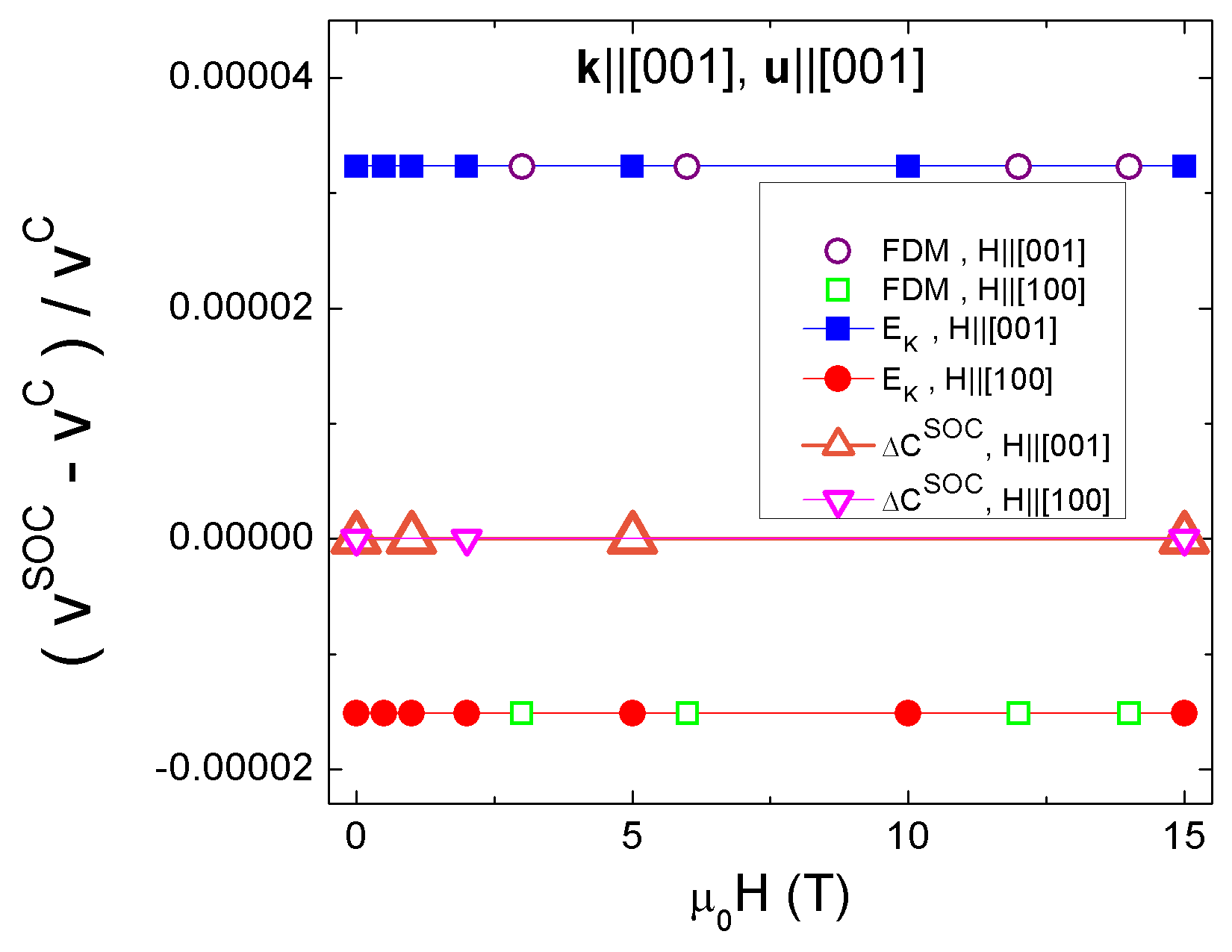}
\caption{Fractional change in velocity of BCC Fe versus the magnitude of the applied magnetic field along $\boldsymbol{H}\parallel[100]$ and $\boldsymbol{H}\parallel[001]$ for a longitudinal wave propagating along $\boldsymbol{k}\parallel[001]$ with polarization $\boldsymbol{u}\parallel[001]$. Open green squares and purple circles stand for the calculations with spin-lattice simulations obtained from the FDM for the cases $\boldsymbol{H}\parallel\boldsymbol{M}\parallel[100]$ and $\boldsymbol{H}\parallel\boldsymbol{M}\parallel[001]$, respectively. Red points and blue squares stand for the calculations with spin-lattice simulations obtained from the oscillation of the kinetic energy for the cases $\boldsymbol{H}\parallel\boldsymbol{M}\parallel[100]$ and $\boldsymbol{H}\parallel\boldsymbol{M}\parallel[001]$, respectively. Open orange and pink triangles represent the calculations using the corrections to the elastic constants given by Eq.\ref{eq:v_k001u001_soc} for the cases $\boldsymbol{H}\parallel\boldsymbol{M}\parallel[100]$ and $\boldsymbol{H}\parallel\boldsymbol{M}\parallel[001]$, respectively.}
\label{fig:hlong}
\end{figure}

Let us now consider a longitudinal wave propagating along $\boldsymbol{k}\parallel[001]$ with polarization $\boldsymbol{u}\parallel[001]$. If the magnetization is along   $\boldsymbol{M}\parallel[001]$, then functions in Eq. \ref{eq:dfani}  read\cite{ROUCHY198069}
\begin{equation}
\begin{aligned}
& G\Bigg\vert_{\boldsymbol{u}\parallel[001]}^{\boldsymbol{k}\parallel[001]\parallel \boldsymbol{M}}  = \frac{2 m_1^{\gamma,2}}{3C^c_{11}},\\ & R\Bigg\vert_{\boldsymbol{u}\parallel[001]}^{\boldsymbol{k}\parallel[001]\parallel \boldsymbol{M}}  = 0,\\
& S\Bigg\vert_{\boldsymbol{u}\parallel[001]}^{\boldsymbol{k}\parallel[001]\parallel\boldsymbol{M}}  =0,\\
& \frac{l^{SOC}-l^c}{l^c}\Bigg\vert^{\boldsymbol{k}\parallel[001]\parallel \boldsymbol{M}}  = \frac{2}{3}\lambda^{\gamma,2}=\lambda_{001},
\label{eq:dfk001u001m001}
\end{aligned}
\end{equation}
while if the magnetization is along $\boldsymbol{M}\parallel[100]$, then we have\cite{ROUCHY198069}
\begin{equation}
\begin{aligned}
 \frac{v^{SOC}-v^{c}}{v^{c}}\Bigg\vert_{\boldsymbol{u}\parallel [001]}^{\boldsymbol{k}\parallel[001], \boldsymbol{M}\parallel[100]} =-\frac{1}{2}\frac{v^{SOC}-v^{c}}{v^{c}}\Bigg\vert_{\boldsymbol{u}\parallel[001]}^{\boldsymbol{k}\parallel\boldsymbol{M}\parallel[001]}.
\label{eq:dfk001u001m100}
\end{aligned}
\end{equation}
We simulated these two cases for different values of the applied magnetic field in Fig.\ref{fig:hlong}. The spin-lattice simulations give a constant fractional change in velocity as a function of the applied field, which is consistent with the lack of field dependency ($S(H)=0$) in Eqs. \ref{eq:dfk001u001m001} and \ref{eq:dfk001u001m100}. We also see that it correctly reproduces the relationship provided by Eq. \ref{eq:dfk001u001m100}, which comes from a high order effect induced by the anisotropic morphic coefficient $m_1^{\gamma,2}$ via the function $G$ and fractional change in length. We can derive the morphic coefficient $m_1^{\gamma,2}$ by fitting the summation of the functions in Eq. \ref{eq:dfk001u001m001}, where the values for $C^c_{11}$ and $\lambda_{001}$ are taken from Table \ref{table:data_properties}, to the fractional change in velocity given by the  spin-lattice simulations in Fig.\ref{fig:hlong}. This procedure gives $m_1^{\gamma,2}=2.4$ MPa using both the FDM and the oscillation of the kinetic energy. The fractional change in velocity derived from the corrections to elastic constants (Eq. \ref{eq:v_k001u001_soc}) is zero for both cases ($\boldsymbol{M}\parallel[001]$ and $\boldsymbol{M}\parallel[100]$) because $\triangle S^{SOC}_{33}(H)=0$, as expected from the fact that $S(H)=0$ in Eq.\ref{eq:dfk001u001m001}. 

\begin{table}[h!]
\caption{Calculated anisotropic morphic coefficients of the spin-lattice model for BCC Fe.}
\label{table:data_ani_morphic}
\centering
\begin{tabular}{ccc}
\toprule
\textbf{Method} 		&
\begin{tabular}[c]{@{}c@{}}\textbf{Morphic} \\ \textbf{ coefficient}\end{tabular}		& \quad\textbf{MPa} \quad\quad \\
\midrule
Kinetic energy & $m_3^{\gamma,2}$  				& -14.3   \\
FDM &  				& -15.3   \\
\midrule
Kinetic energy & $m_1^{\gamma,2}$                      & 2.4  \\
FDM & 					& 2.4  \\
\bottomrule
\end{tabular}
\end{table}

In Table \ref{table:data_ani_morphic}, we present a summary of the two anisotropic morphic coefficients derived for the spin-lattice model of BCC Fe in this work.  There are in total 9 morphic coefficients $m^{\mu,2}_i$ (where $\mu=\alpha,\beta,\gamma$ and $i=1,2,3$) up to second order in the direction cosine of magnetization, see Appendix \ref{app:Internal_energy}\cite{Rouchy1979,ROUCHY198069,booktremolet}. Although some of these morphic coefficients were not computed in this study, they can be evaluated in a similar way by choosing other propagating directions of the wave\cite{ROUCHY198069}. 
In Tables \ref{table:data_iso_morphic} and \ref{table:data_ani_morphic} we see that the isotropic morphic coefficient $m_{3}^{\alpha,2}$ (induced by the exchange interaction) is about four orders of magnitude larger than the anisotropic ones.
This result is in good agreement with morphic coefficients reported for FCC Ni, where similar differences were observed~\cite{booktremolet,DUTREMOLETDELACHEISSERIE198277}.  
Note that the extrapolation of the calculated morphic coefficients with this spin-lattice model to experiment is not obvious because they are linear combinations of third order in strain elastic constants ($\tilde{C}_{ijk}$) and second order in strain MEL constants ($\tilde{M}^{\mu,2}_i$) \cite{Rouchy1979,ROUCHY198069,booktremolet}. The SNAP interatomic potential used in the model might  describe,  at least to some extent,  experimental high order elastic constants  thanks to its quantum mechanical derivation and complex functional form~\cite{Nikolov2021,thompson2015spectral}. However, it is not clear that the  N\'{e}el model used in this work could be sufficiently accurate to describe correctly experimental high order MEL constants $\tilde{M}^{\mu,2}_i$ since it was originally designed to reproduce the experimental  MEL constants $B^{\mu,2}$ only up to first order in strain\cite{nieves2021spinlattice_prb}. Possible deviations due to the N\'{e}el model might not be relevant in the calculation of those morphic coefficients where the contribution of high order elastic constants is much greater than the high order MEL constants ($\tilde{M}^{\mu,2}_i\ll\tilde{C}_{ljk}\lambda^{\mu,2}$, see Eq.\ref{eq:morphic_coeff})\cite{ROUCHY198069}.

\begin{figure}[h]
\centering
\includegraphics[width=\columnwidth ,angle=0]{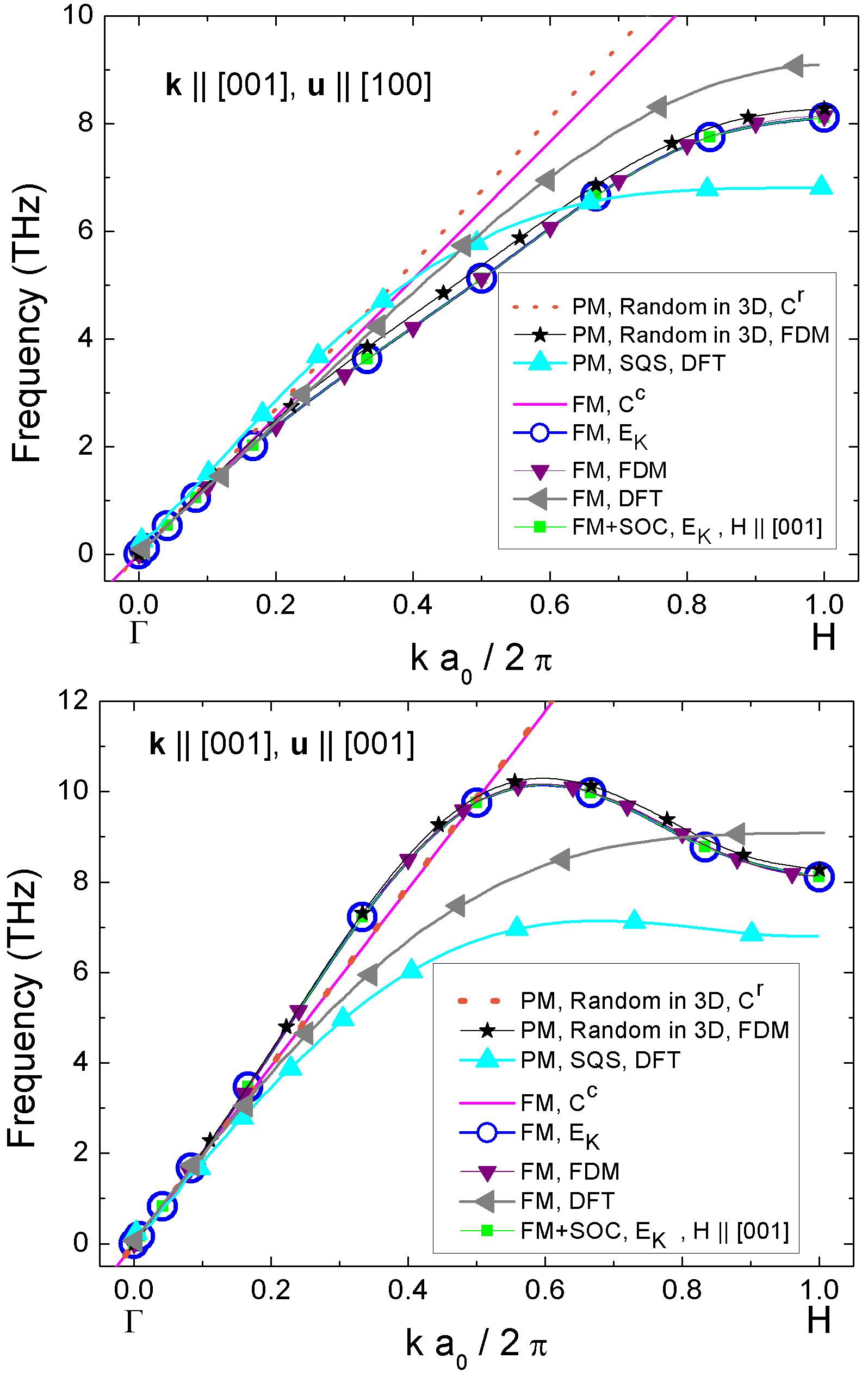}
\caption{Phonon dispersion of (top) transverse and (bottom) longitudinal modes. Black stars correspond to calculations with the FDM at paramagnetic (PM) state with randomly oriented magnetic moments in 3D. Dash orange and solid pink lines stand for the calculations with the elastic constants at the PM and ferromagnetic collinear without SOC states (FM), respectively. Blue circles and violet triangles show data from the oscillation of the kinetic energy and FDM at a collinear state without SOC, respectively. Green squares represent calculations obtained from the oscillation of the kinetic energy at a collinear state with SOC under an applied magnetic field $\mu_0\boldsymbol{H}=(0,0,40)$ T. Cyan and grey triangles give DFT calculations performed by Ikeda et al.\cite{Ikeda} using SQS for the PM and FM states, respectively.}
\label{fig:disp}
\end{figure}

\subsection{Magnetoelastic effects on phonon dispersion}
\label{subsection:dispersion}

\begin{figure}[h!]
\centering
\includegraphics[width=\columnwidth ,angle=0]{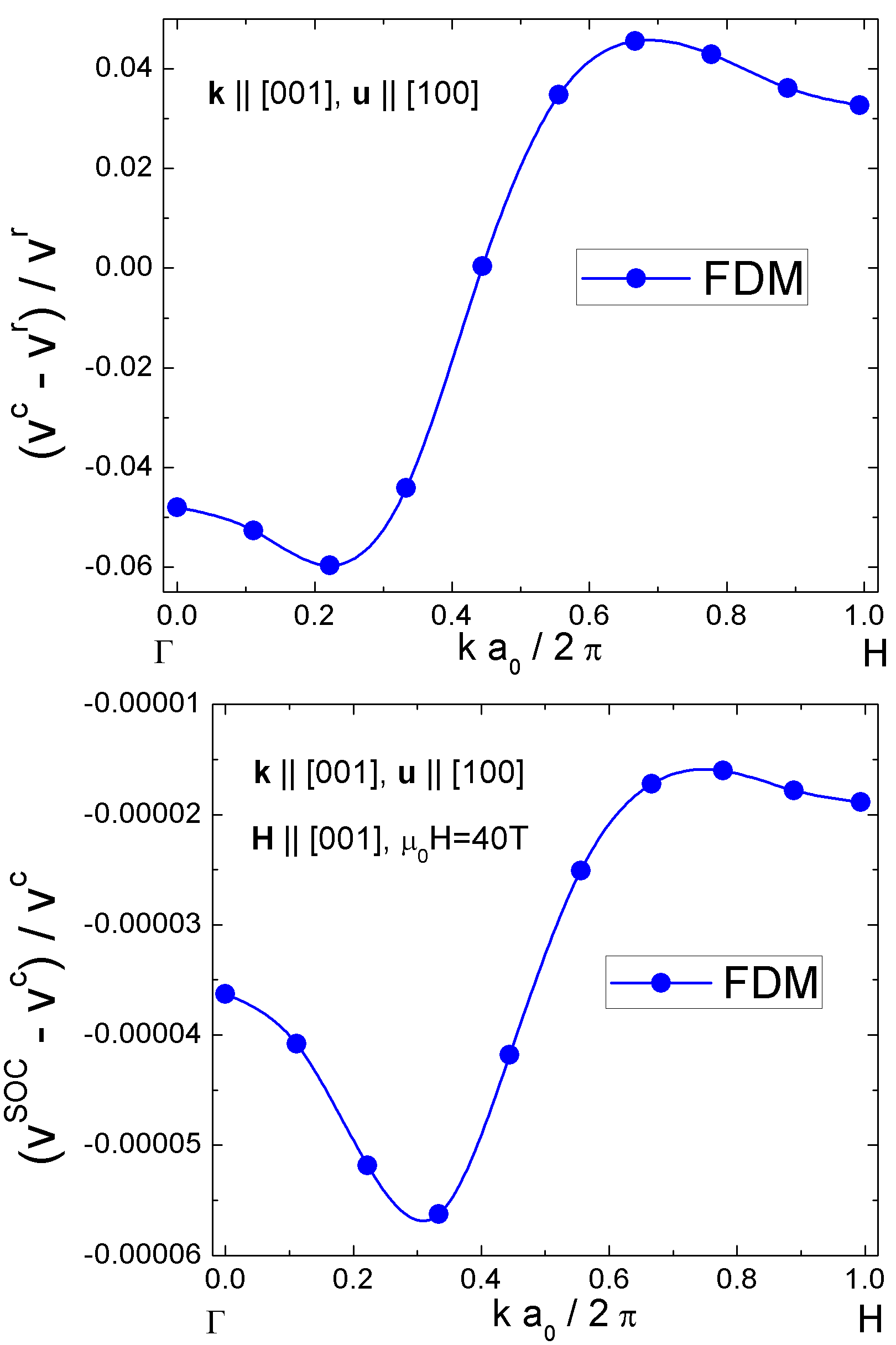}
\caption{Fractional change in group velocity of phonons versus phonon's momentum along $\boldsymbol{k}\parallel[001]$ with polarization $\boldsymbol{u}\parallel[100]$. (Top) Fractional change in velocity of the collinear state (including only exchange interaction) with respect to the paramagnetic state with randomly oriented magnetic moments in 3D. (Bottom) Fractional change in velocity of the collinear state (including both exchange interaction and SOC) with respect to collinear state including only exchange interaction. The calculations are performed using the FDM.}
\label{fig:dv_k}
\end{figure}

In previous sections, we studied the MEL effects on phonons with low momentum ($k_{ph}\rightarrow 0$) in order to compare the atomistic simulations with the continuum theory of sound velocity in a solid. Here, we extend our study by calculating the MEL effects on the frequency of phonons with larger momentum. In particular, we compute the phonon dispersion along the $\Gamma-$H line of k-points, see Fig.\ref{fig:disp}. In general, we observe small changes in phonon's frequency due to MEL effects. For comparison, we also plot in Fig.\ref{fig:disp} the Density-Functional Theory (DFT) calculations performed by Ikeda et al.\cite{Ikeda} using SQS with size $2\times2\times2$. Close to the $\Gamma$ point ($k_{ph}\rightarrow 0$), our calculations with the spin-lattice model are very similar to the DFT results reported in Ref.\cite{Ikeda}. However, at higher values of $k$, we obtain that the phonon frequencies of the paramagnetic state are slightly larger than in the ferromagnetic state, while the opposite behaviour was found by Ikeda et al.\cite{Ikeda} with also a higher shift in the frequencies. Similar results as in Ref.\cite{Ikeda} have been reported by Novikov et al.\cite{novikov2021magnetic} using DFT and magnetic moment tensor potentials. Finally, in Fig.\ref{fig:dv_k} we analyze how the fractional change in phonon group velocity is influenced by the magnitude of phonon momentum $k_{ph}$. We observe a significant modification of $(v^c-v^r)/v^r$ as we increase $k_{ph}$, where the sign is changed in the middle of $\Gamma$-H line. This result suggests that the theoretical equation derived for sound waves Eq.\ref{eq:dfk001u100m001iso} might only hold for phonon with low momentum ($k_{ph}\rightarrow 0$). On the other hand, $(v^{SOC}-v^c)/v^c$ is not so strongly affected by the magnitude of phonon momentum, which means that in some cases the theoretical fractional change in velocity derived for sound waves might still provide at least a reasonable qualitative description for phonons with larger momentum.

\section{Conclusions}
\label{section:conclusions}

In summary, through the analysis of the kinetic energy and forces, we found that the spin-lattice model for BCC Fe is capable to describe the four main MEL effects
on sound velocity (isotropic exchange
effects,  anisotropic morphic effects,  field dependent
effects  and  rotational-magnetostrictive effects). We also showed that the alternative approach based on corrections to the elastic constants  can correctly account for isotropic exchange effects and anisotropic field dependent effects, but not for high order anisotropic effects (like anisotropic morphic effects and rotational-magnetostrictive effects) because the used SOC corrections were derived from the linear MEL theory\cite{Rinaldi1985}. In analogy to experiment, some morphic coefficients of this spin-lattice model were extracted from the analysis of the fractional change in frequency, exhibiting a similar order of magnitude as in FCC Ni\cite{DUTREMOLETDELACHEISSERIE198277,booktremolet}. In the low magnetic field regime, our simulations reveal an interesting non-monotonic dependence on the applied field of the fractional change in sound velocity, which cannot be described by the current theory since it is valid only at high applied fields\cite{Simon+1958+84+89}. Hence, the presented computational framework could be a useful tool to explore and understand the field regimes still uncovered by the theory. Similarly, this approach could also be helpful to study the combination of magnetic and atomistic scale effects (like vacancies, impurities or complex defects) on sound velocity, as well as design novel technological magneto-acoustic applications.

\section*{Acknowledgement}

This work was supported by the ERDF in the IT4Innovations national supercomputing center - path to exascale project (CZ.02.1.01/0.0/0.0/16-013/0001791) within the OPRDE and projects “e-INFRA CZ (ID:90140)" and Donau No. 8X20050 by The Ministry of Education, Youth and Sports of the Czech Republic. In addition, P.N. and D.L. acknowledge support from the H2020-FETOPEN no.~863155 s-NEBULA project. S. Nikolov is an employee of Sandia National Laboratories, a multimission laboratory managed and operated by National Technology and Engineering Solutions of Sandia, LLC, a wholly owned subsidiary of Honeywell International Inc., for the U.S. Department of Energy’s National Nuclear Security Administration under Contract DE-NA0003525. This paper describes objective technical results and analysis. Any subjective views or opinions that might be expressed in the paper do not necessarily represent the views of the U.S. Department of Energy or the United States Government.

\appendix

\section{Magnetoelastic effects on sound velocity as corrections to the elastic constants}
\label{app:eff_cij}

MEL effects on sound velocity can be described in terms of effective elastic constants. In this appendix, we present a brief review of this approach. Let us start by deriving the general expression of sound velocity in terms of the elastic constants. The equation of motion of an elastic wave is given by\cite{Landau}
\begin{equation}
\begin{aligned}
     \rho\frac{\partial^2 u_i}{\partial t^2}=\sum_{j=x,y,z}\frac{\partial\sigma_{ij}}{\partial r_j},\quad\quad i=x,y,z
    \label{eq:eq_mot}
\end{aligned}
\end{equation}
where $ \rho$ is the mass density, $\boldsymbol{u}$ is the displacement vector and $\sigma_{ij}$ is the stress tensor that is related to the fourth-order elastic stiffness tensor $c_{ijkl}$ and the second-order strain tensor $\epsilon_{ij}$ through the generalized Hooke's law
\begin{equation}
    \sigma_{ij} = \sum_{k,l=x,y,z}c_{ijkl}\epsilon_{kl}, \quad i,j=x,y,z.
    \label{eq:hooke}
\end{equation}
Using the symmetry of the stress and strain tensors, the Hooke's law can be written in matrix notation as
\begin{equation}
\begin{aligned}
    \begin{pmatrix}
\sigma_{xx} \\
\sigma_{yy}   \\
\sigma_{zz}   \\
\sigma_{yz}   \\
\sigma_{zx}   \\
\sigma_{xy}   \\
\end{pmatrix} & =\begin{pmatrix}
c_{xxxx} & c_{xxyy} & c_{xxzz} & c_{xxyz} & c_{xxzx} & c_{xxxy} \\
c_{yyxx} & c_{yyyy} & c_{yyzz} & c_{yyyz} & c_{yyzx} & c_{yyxy} \\
c_{zzxx} & c_{zzyy} & c_{zzzz} & c_{zzyz} & c_{zzzx} & c_{zzxy} \\
c_{yzxx} & c_{yzyy} & c_{yzzz} & c_{yzyz} & c_{yzzx} & c_{yzxy} \\
c_{zxxx} & c_{zxyy} & c_{zxzz} & c_{zxyz} & c_{zxzx} & c_{zxxy} \\
c_{xyxx} & c_{xyyy} & c_{xyzz} & c_{xyyz} & c_{xyzx} & c_{xyxy} \\
\end{pmatrix}
    \begin{pmatrix}
\epsilon_{xx} \\
\epsilon_{yy}   \\
\epsilon_{zz}   \\
2\epsilon_{yz}\\
2\epsilon_{zx} \\
2\epsilon_{xy} \\
\end{pmatrix}
\label{eq:stiffnees_tensor}
\end{aligned}
\end{equation}
To facilitate the manipulation of this equation it is convenient to define the following six-dimensional vectors (Voigt notation)
\begin{equation}
\begin{aligned}
  \boldsymbol{\tilde{\sigma}} =\begin{pmatrix}
\tilde{\sigma}_{1} \\
\tilde{\sigma}_{2}   \\
\tilde{\sigma}_{3}   \\
\tilde{\sigma}_{4}   \\
\tilde{\sigma}_{5}   \\
\tilde{\sigma}_{6}   \\
\end{pmatrix} 
=\begin{pmatrix}
\sigma_{xx} \\
\sigma_{yy}   \\
\sigma_{zz}   \\
\sigma_{yz}   \\
\sigma_{zx}   \\
\sigma_{xy}   \\
\end{pmatrix}, \quad\quad
 \boldsymbol{\tilde{\epsilon}} =\begin{pmatrix}
\tilde{\epsilon}_{1} \\
\tilde{\epsilon}_{2}   \\
\tilde{\epsilon}_{3}   \\
\tilde{\epsilon}_{4}   \\
\tilde{\epsilon}_{5}   \\
\tilde{\epsilon}_{6}   \\
\end{pmatrix} 
=\begin{pmatrix}
\epsilon_{xx} \\
\epsilon_{yy}   \\
\epsilon_{zz}   \\
2\epsilon_{yz}   \\
2\epsilon_{zx}   \\
2\epsilon_{xy}   \\
\end{pmatrix},
\label{eq:vector_stress_strain}
\end{aligned}
\end{equation}
and replace $c_{ijkl}$ by $C_{nm}$ contracting a pair of cartesian indices into a single integer: $xx\rightarrow1$, $yy\rightarrow2$, $zz\rightarrow3$, $yz\rightarrow4$, $zx\rightarrow5$ and $xy\rightarrow6$. Using these conversion rules the Hooke's law is simplified to 
\begin{equation}
    \tilde{\sigma}_{i} = \sum_{j=1}^6 C_{ij}\tilde{\epsilon}_{j}, \quad i=1,...,6
    \label{eq:hooke_simple}
\end{equation}
where in matrix form reads
\begin{equation}
\begin{aligned}
    \begin{pmatrix}
\tilde{\sigma}_{1} \\
\tilde{\sigma}_{2}   \\
\tilde{\sigma}_{3}   \\
\tilde{\sigma}_{4}   \\
\tilde{\sigma}_{5}   \\
\tilde{\sigma}_{6}   \\
\end{pmatrix} & =\begin{pmatrix}
C_{11} & C_{12} & C_{13} & C_{14} & C_{15} & C_{16} \\
C_{21} & C_{22} & C_{23} & C_{24} & C_{25} & C_{26} \\
C_{31} & C_{32} & C_{33} & C_{34} & C_{35} & C_{36} \\
C_{41} & C_{42} & C_{43} & C_{44} & C_{45} & C_{46} \\
C_{51} & C_{52} & C_{53} & C_{54} & C_{55} & C_{56} \\
C_{61} & C_{62} & C_{63} & C_{64} & C_{65} & C_{66} \\
\end{pmatrix}
    \begin{pmatrix}
\tilde{\epsilon}_{1} \\
\tilde{\epsilon}_{2}   \\
\tilde{\epsilon}_{3}   \\
\tilde{\epsilon}_{4}   \\
\tilde{\epsilon}_{5}   \\
\tilde{\epsilon}_{6}   \\
\end{pmatrix}.
\label{eq:stiffnees_tensor_voigt}
\end{aligned}
\end{equation}
We additionally assume that our material is hyperelastic (there exists an elastic energy function) what implies that  $C_{ij}=C_{ji}$. Next, using Eqs.\ref{eq:vector_stress_strain} and \ref{eq:hooke_simple} in the equation of motion Eq.\ref{eq:eq_mot} gives
\begin{equation}
\begin{aligned}
     \frac{\partial^2 u_x}{\partial t^2}=\sum_{i=1}^6\left(C_{1i}\frac{\partial}{\partial r_x}+C_{6i}\frac{\partial}{\partial r_y}+C_{5i}\frac{\partial}{\partial r_z}\right)\tilde{\epsilon}_{i},\\
     \frac{\partial^2 u_y}{\partial t^2}=\sum_{i=1}^6\left(C_{6i}\frac{\partial}{\partial r_x}+C_{2i}\frac{\partial}{\partial r_y}+C_{4i}\frac{\partial}{\partial r_z}\right)\tilde{\epsilon}_{i},\\
     \frac{\partial^2 u_z}{\partial t^2}=\sum_{i=1}^6\left(C_{5i}\frac{\partial}{\partial r_x}+C_{4i}\frac{\partial}{\partial r_y}+C_{3i}\frac{\partial}{\partial r_z}\right)\tilde{\epsilon}_{i}.
    \label{eq:eq_mot2}
\end{aligned}
\end{equation}
For small deformations (infinitesimal strain theory), the strain tensor can be expressed in terms of the displacement vector $\boldsymbol{u}$ as\cite{Landau}
\begin{equation}
\begin{aligned}
     \epsilon_{ij}=\frac{1}{2}\left(\frac{\partial u_{i}}{\partial r_{j}}+\frac{\partial u_j}{\partial r_i}\right),\quad\quad i,j=x,y,z
    \label{eq:disp_vec}
\end{aligned}
\end{equation}
Hence, combining this equation with Eq.\ref{eq:vector_stress_strain}, the six-dimensional vector $\tilde{\epsilon}_{i}$ can be expressed in terms of the displacement vector as 
\begin{equation}
\begin{aligned}
 \boldsymbol{\tilde{\epsilon}} =\begin{pmatrix}
\tilde{\epsilon}_{1} \\
\tilde{\epsilon}_{2}   \\
\tilde{\epsilon}_{3}   \\
\tilde{\epsilon}_{4}   \\
\tilde{\epsilon}_{5}   \\
\tilde{\epsilon}_{6}   \\
\end{pmatrix} 
=\begin{pmatrix}
\epsilon_{xx} \\
\epsilon_{yy}   \\
\epsilon_{zz}   \\
2\epsilon_{yz}   \\
2\epsilon_{zx}   \\
2\epsilon_{xy}   \\
\end{pmatrix}
=\begin{pmatrix}
\frac{\partial u_{x}}{\partial r_{x}} \\
\frac{\partial u_{y}}{\partial r_{y}}   \\
\frac{\partial u_{z}}{\partial r_{z}}   \\
\frac{\partial u_{y}}{\partial r_{z}}+\frac{\partial u_z}{\partial r_y}  \\
\frac{\partial u_{x}}{\partial r_{z}}+\frac{\partial u_z}{\partial r_x}   \\
\frac{\partial u_{x}}{\partial r_{y}}+\frac{\partial u_y}{\partial r_x}   \\
\end{pmatrix}.
\label{eq:vector_stress_strain2}
\end{aligned}
\end{equation}
Last, replacing Eq.\ref{eq:vector_stress_strain2} in Eq.\ref{eq:eq_mot2} and considering a monochromatic elastic wave $u_i=u_{0,i}e^{i(\boldsymbol{k}\cdot\boldsymbol{r}+2\pi f t)}$, we find\cite{Landau}
\begin{equation}
\begin{aligned}
\begin{pmatrix}
A_{xx} & A_{xy} & A_{xz}  \\
A_{yx} & A_{yy} & A_{yz} \\
A_{zx} & A_{zy} & A_{zz} \\
\end{pmatrix}\cdot
    \begin{pmatrix}
u_{x} \\
u_{y}   \\
u_{z}   \\
\end{pmatrix}
=\begin{pmatrix}
0 \\
0   \\
0   \\
\end{pmatrix}
\label{eq:matrix_product}
\end{aligned}
\end{equation}
where
\begin{equation}
\begin{aligned}
     A_{xx} & = -\rho (2\pi f)^2 + (k_x C_{11} + k_y C_{61} + k_z C_{51}) k_x\\
   & + (C_{15} k_x + C_{65} k_y + C_{55} k_z) k_z  + (k_x C_{16} + k_y C_{66} + k_z C_{56}) k_y,\\
A_{xy} & = (C_{14} k_x + C_{64} k_y + C_{54} k_z) k_z + (k_x C_{12} + k_y C_{62} + 
     k_z C_{52}) k_y \\
     & + (C_{16} k_x + C_{66} k_y + C_{56} k_z) k_x,\\
A_{xz} & = (C_{14} k_x + C_{64} k_y + C_{54} k_z) k_y + (C_{15} k_x + C_{65} k_y + C_{55} k_z) k_x\\
     & + (C_{13} k_x + C_{63} k_y + C_{53} k_z) k_z,\\
A_{yx} & = (k_x C_{61} + k_y C_{21} + k_z C_{41}) k_x + (C_{65} k_x + C_{25} k_y + 
     C_{45} k_z) k_z \\
     & + (k_x C_{66} + k_y C_{26} + k_z C_{46}) k_y,\\
A_{yy} & = -\rho  (2\pi f)^2 + (C_{64} k_x + C_{24} k_y + C_{44} k_z) k_z\\
   & + (k_x C_{62} + k_y C_{22} + 
     k_z C_{42}) k_y + (C_{66} k_x + C_{26} k_y + C_{46} k_z) k_x,\\
A_{yz} & = (C_{64} k_x + C_{24} k_y + C_{44} k_z) k_y + (C_{65} k_x + C_{25} k_y  + C_{45} k_z) k_x \\
& + (C_{63} k_x + C_{23} k_y + C_{43} k_z) k_z,\\
A_{zx} & = (k_x C_{51} + k_y C_{41} + k_z C_{31}) k_x + (C_{55} k_x + C_{45} k_y + 
     C_{35} k_z) k_z\\
     & + (k_x C_{56} + k_y C_{46} + k_z C_{36}) k_y,\\
A_{zy} & = (C_{54} k_x + C_{44} k_y + C_{34} k_z) k_z + (k_x C_{52} + k_y C_{42} + 
     k_z C_{32}) k_y \\
     & + (C_{56} k_x + C_{46} k_y + C_{36} k_z) k_x,\\
A_{zz} & = -\rho (2\pi f)^2 + (C_{54} k_x + C_{44} k_y + C_{34} k_z) k_y \\
   & + (C_{55} k_x + C_{45} k_y + 
     C_{35} k_z) k_x + (C_{53} k_x + C_{43} k_y + C_{33} k_z) k_z.
    \label{eq:Aij}
\end{aligned}
\end{equation}
From Eq.\ref{eq:matrix_product} one can straightforwardly derive the sound velocity. For example, in the case of a transverse wave propagating along $\boldsymbol{k}\parallel[001]$ ($k_x=0$ and $k_y=0$) with polarization $\boldsymbol{u}\parallel[100]$ ($u_y=0$ and $u_z=0$) we obtain the following relation from Eq.\ref{eq:matrix_product}
\begin{equation}
\begin{aligned}
     \left[ -\rho (2\pi f)^2 +C_{55}k_z^2\right]u_x=0,
    \label{eq:eq_trans}
\end{aligned}
\end{equation}
hence in this case the frequency $f$ is related to the wave vector $\boldsymbol{k}$ as
\begin{equation}
\begin{aligned}
     2\pi f(\boldsymbol{k})\Bigg\vert_{\boldsymbol{u}\parallel[100]}^{\boldsymbol{k}\parallel[001]} = k_z \sqrt{\frac{C_{55}}{\rho}}.
    \label{eq:eq_trans2}
\end{aligned}
\end{equation}
The velocity of propagation of the wave (group velocity) $\boldsymbol{v}$ is given by the derivative of the frequency $2\pi f$ with respect to the wave vector $\boldsymbol{k}$\cite{Landau} 
\begin{equation}
\begin{aligned}
     \boldsymbol{v}=\frac{\partial (2\pi f)}{\partial \boldsymbol{k}}=\left(\frac{\partial (2\pi f)}{\partial k_x},\frac{\partial (2\pi f)}{\partial k_y},\frac{\partial (2\pi f)}{\partial k_z}\right).
    \label{eq:eq_trans3}
\end{aligned}
\end{equation}
Applying Eq.\ref{eq:eq_trans3} to Eq. \ref{eq:eq_trans2} yields
\begin{equation}
\begin{aligned}
     \boldsymbol{v}\Bigg\vert_{\boldsymbol{u}\parallel[100]}^{\boldsymbol{k}\parallel[001]}=\frac{\partial (2\pi f)}{\partial \boldsymbol{k}}\Bigg\vert_{\boldsymbol{u}\parallel[100]}^{\boldsymbol{k}\parallel[001]}=\left(0,0,\sqrt{\frac{C_{55}}{\rho}}\right),
    \label{eq:v_k001u100_vec}
\end{aligned}
\end{equation}
thus the magnitude of sound velocity is
\begin{equation}
\begin{aligned}
     v\Bigg\vert_{\boldsymbol{u}\parallel[100]}^{\boldsymbol{k}\parallel[001]}=\sqrt{\frac{C_{55}}{\rho}}.
    \label{eq:v_k001u100}
\end{aligned}
\end{equation}
Similarly, for the longitudinal mode $\boldsymbol{u}\parallel[001]$ ($u_x=0$ and $u_y=0$) we obtain
\begin{equation}
\begin{aligned}
     v\Bigg\vert_{\boldsymbol{u}\parallel[001]}^{\boldsymbol{k}\parallel[001]}=\sqrt{\frac{C_{33}}{\rho}}.
    \label{eq:v_k001u001}
\end{aligned}
\end{equation}
Note that for these particular cases the sound velocity is parallel to the wave vector ($\boldsymbol{v}\parallel\boldsymbol{k}$), however this relationship do not generally hold in crystals\cite{Landau}. It is generally true in an isotropic body\cite{Landau}. 

Next, for cubic crystals we may write the effective elastic tensor that includes all MEL effects as
\begin{equation}
\begin{aligned}
     C_{ij}=C_{ij}^r+\triangle C_{ij}^J+\triangle C_{ij}^{SOC},
    \label{eq:C_eff}
\end{aligned}
\end{equation}
where
\begin{equation}
\begin{aligned}
& C^r  =\begin{pmatrix}
C^r_{11} & C^r_{12} & C^r_{12} & 0 & 0 & 0 \\
C^r_{12} & C^r_{11} & C^r_{12} & 0 & 0 & 0 \\
C^r_{12} & C^r_{12} & C^r_{11} & 0 & 0 & 0 \\
0 & 0 & 0 & C^r_{44} & 0 & 0 \\
0 & 0 & 0 & 0 &  C^r_{44} & 0 \\
0 & 0 & 0 & 0 & 0 &  C^r_{44} \\
\end{pmatrix},\\
&  \triangle C^J  =  \begin{pmatrix}
\triangle C^J_{11} & \triangle C^J_{12} & \triangle C^J_{12} & 0 & 0 & 0 \\
\triangle C^J_{12} & \triangle C^J_{11} & \triangle C^J_{12} & 0 & 0 & 0 \\
\triangle C^J_{12} & \triangle C^J_{12} & \triangle C^J_{11} & 0 & 0 & 0 \\
0 & 0 & 0 & \triangle C^J_{44} & 0 & 0 \\
0 & 0 & 0 & 0 &  \triangle C^J_{44} & 0 \\
0 & 0 & 0 & 0 & 0 &  \triangle C^J_{44} \\
\end{pmatrix},\\
& \triangle C^{SOC}  =\\
 & \begin{pmatrix}
 \triangle C^{SOC}_{11} & \triangle C^{SOC}_{12} & \triangle C^{SOC}_{13} & \triangle C^{SOC}_{14} & \triangle C^{SOC}_{15} & \triangle C^{SOC}_{16} \\
\triangle C^{SOC}_{12} & \triangle C^{SOC}_{22} & \triangle C^{SOC}_{23} & \triangle C^{SOC}_{24} & \triangle C^{SOC}_{25} & \triangle C^{SOC}_{26} \\
\triangle C^{SOC}_{13} & \triangle C^{SOC}_{23} & \triangle C^{SOC}_{33} & \triangle C^{SOC}_{34} & \triangle C^{SOC}_{35} & \triangle C^{SOC}_{36} \\
\triangle C^{SOC}_{14} & \triangle C^{SOC}_{24} & \triangle C^{SOC}_{34} & \triangle C^{SOC}_{44} & \triangle C^{SOC}_{45} & \triangle C^{SOC}_{46} \\
\triangle C^{SOC}_{15} & \triangle C^{SOC}_{25} & \triangle C^{SOC}_{35} & \triangle C^{SOC}_{45} & \triangle C^{SOC}_{55} & \triangle C^{SOC}_{56} \\
\triangle C^{SOC}_{16} & \triangle C^{SOC}_{26} & \triangle C^{SOC}_{36} & \triangle C^{SOC}_{46} & \triangle C^{SOC}_{56} & \triangle C^{SOC}_{66} \\
\end{pmatrix}.\\
\label{eq:C_eff2}
\end{aligned}
\end{equation}
The term $C^r$ corresponds to elastic tensor at the paramagnetic-like state (i.e. with randomly oriented atomic magnetic moments). The term $\triangle C^J$ gives the correction to the elastic tensor when the isotropic exchange interaction is included and the system is at the collinear state. Hence, we have
\begin{equation}
\begin{aligned}
     \triangle C_{11}^J & = C^c_{11}-C^r_{11},\\
     \triangle C_{12}^J & = C^c_{12}-C^r_{12},\\
     \triangle C_{44}^J & = C^c_{44}-C^r_{44},
    \label{eq:C_eff_J}
\end{aligned}
\end{equation}
where $C^c_{ij}$ is the elastic tensor when the isotropic exchange interaction is included and the system is at the collinear state. Note that including the isotropic exchange interaction does not change the cubic crystal symmetry, so that  $\triangle C^J$ has the same symmetry as $C^r$. The term $\triangle C^{SOC}$ provides the correction to the elastic tensor when SOC is included\cite{Rinaldi1985}, and depends on the applied magnetic field $\triangle C^{SOC}(\boldsymbol{H})$. The correction due to SOC can lower the crystal symmetry, so that the symmetries of $\triangle C^{SOC}$ could be different to $C^c$ and  $\triangle C^J$.

In the case of a transverse wave propagating along $\boldsymbol{k}\parallel[001]$ with polarization $\boldsymbol{u}\parallel[100]$, combining Eqs.\ref{eq:v_k001u100} and \ref{eq:C_eff}, we find that the fractional change in velocity  when the system changes from a paramagnetic state ($C_{55}=C^r_{44}$) to a collinear state including only the isotropic exchange ($C_{55}=C^r_{44}+\triangle C^J_{44}=C^c_{44}$) is
\begin{equation}
\begin{aligned}
     \frac{v^c-v^r}{v^r}\Bigg\vert_{\boldsymbol{u}\parallel[100]}^{\boldsymbol{k}\parallel[001]}=\frac{\sqrt{\frac{C^c_{44}}{\rho^c}}-\sqrt{\frac{C^r_{44}}{\rho^r}}}{\sqrt{\frac{C^r_{44}}{\rho^r}}}.
    \label{eq:v_k001u100_para}
\end{aligned}
\end{equation}
Similarly, if we include both the isotropic exchange and SOC ($C_{55}=C^r_{44}+\triangle C^J_{44}+\triangle C^{SOC}_{55}=C^c_{44}+\triangle C^{SOC}_{55}$), then 
 the fractional change in velocity with respect to the collinear state with only isotropic exchange reads
\begin{equation}
\begin{aligned}
     \frac{v^{SOC}-v^c}{v^c}\Bigg\vert_{\boldsymbol{u}\parallel[100]}^{\boldsymbol{k}\parallel[001]}=\frac{\sqrt{\frac{C^c_{44}+\triangle C_{55}^{SOC}}{\rho^{SOC}}}-\sqrt{\frac{C^c_{44}}{\rho^c}}}{\sqrt{\frac{C^c_{44}}{\rho^c}}}.
    \label{eq:v_k001u100_soc}
\end{aligned}
\end{equation}
In the case of the longitudinal mode, using Eq.\ref{eq:v_k001u001}, we obtain

\begin{eqnarray}
     \frac{v^c-v^r}{v^r}\Bigg\vert_{\boldsymbol{u}\parallel[001]}^{\boldsymbol{k}\parallel[001]} &  = & \frac{\sqrt{\frac{C^c_{11}}{\rho^c}}-\sqrt{\frac{C^r_{11}}{\rho^r}}}{\sqrt{\frac{C^r_{11}}{\rho^r}}},\label{eq:v_k001u001_soc0}\\
     \frac{v^{SOC}-v^c}{v^c}\Bigg\vert_{\boldsymbol{u}\parallel[001]}^{\boldsymbol{k}\parallel[001]} & = & \frac{\sqrt{\frac{C^c_{11}+\triangle C_{33}^{SOC}}{\rho^{SOC}}}-\sqrt{\frac{C^c_{11}}{\rho^{c}}}}{\sqrt{\frac{C^c_{11}}{\rho^{c}}}}.
    \label{eq:v_k001u001_soc}
\end{eqnarray}

In this work, we use the elements of the tensor $\triangle C^{SOC}_{ij}$  calculated by Rinaldi and Turilli\cite{Rinaldi1985} for cubic crystals based on the linear MEL theory. Hence, these elements can not describe the high order effect coming from morphic coefficients (function $G(m)$ in Eq.\ref{eq:dfani})). Similarly, they can not account for the rotational-magnetostrictive effect (function $R(\lambda)$ in Eq.\ref{eq:dfani})) because it requires the finite strain theory\cite{Rouchy1979}. Consequently, the tensor $\triangle C^{SOC}_{ij}$  calculated by Rinaldi and Turilli\cite{Rinaldi1985} can only describe the Simon effect\cite{Simon+1958+84+89}, that is, the field dependent term $S(H)$ in Eq.\ref{eq:dfani} \cite{booktremolet}. The elements $\triangle C^{SOC}_{55}$ and $\triangle C^{SOC}_{33}$ in Eqs.\ref{eq:v_k001u100_soc} and \ref{eq:v_k001u001_soc} calculated by Rinaldi and Turilli read\cite{Rinaldi1985}

\begin{equation}
\begin{aligned}
     \triangle C^{SOC}_{55} & =-\frac{b_2^2}{M_S^2}\left[ (\alpha^0_z)^2\chi_{xx} + (\alpha_x^0)^2\chi_{zz} +2 \alpha_x^0\alpha_z^0\chi_{xz}\right],\\
     \triangle C^{SOC}_{33} & = -\frac{4b_1^2 (\alpha^0_z)^2\chi_{zz}}{M_S^2},
    \label{eq:dc55}
\end{aligned}
\end{equation}
where

\begin{equation}
\begin{aligned}
     \chi_{xx}& =M_S^2\left[\frac{\cos^2\theta_0\cos^2\varphi_0}{E_{\theta\theta}}+\frac{\sin^2\theta_0\sin^2\varphi_0}{E_{\varphi\varphi}}\right],\\
     \chi_{xz} & = -M_S^2\frac{\sin\theta_0\cos\theta_0\cos\varphi_0}{E_{\theta\theta}} ,\\
     \chi_{zz} & = M_S^2\frac{\sin^2\theta_0}{E_{\theta\theta}}.
    \label{eq:chi}
\end{aligned}
\end{equation}

The quantities $\alpha_x^0=\sin\theta_0\cos\varphi_0$, $\alpha_y^0=\sin\theta_0\sin\varphi_0$ and $\alpha_z^0=\cos\theta_0$ are the equilibrium direction cosine of magnetization that minimizes the magnetic energy $E$ given by

\begin{equation}
\begin{aligned}
     E(\theta,\varphi) & =\left(K_1+\frac{b_1^2}{C^c_{11} - C^c_{12}} -\frac{b_2^2}{2C^c_{44}}\right)(\sin^2\theta \cos^2\varphi\sin^2\theta\sin^2\varphi\\
     & + \sin^2\theta\cos^2\varphi \cos^2\theta + \sin^2\theta \sin^2\varphi \cos^2\theta)\\
     & - \mu_0 M_s(H_x\sin\theta \cos\varphi + H_y\sin\theta \sin\varphi + H_z\cos\theta),
    \label{eq:E_mag}
\end{aligned}
\end{equation}

where

\begin{equation}
\begin{aligned}
     E_{\theta\theta}& = \frac{1}{2}\frac{\partial^2 E}{\partial \theta^2}\Bigg\vert_{\theta=\theta_0,\varphi=\varphi_0} ,\\
     E_{\varphi\varphi}& = \frac{1}{2}\frac{\partial^2 E}{\partial \varphi^2}\Bigg\vert_{\theta=\theta_0,\varphi=\varphi_0}.
    \label{eq:dE_mag}
\end{aligned}
\end{equation}

\section{Internal energy in the Lagrangian description}
\label{app:Internal_energy}

It is convenient to work out the internal energy including high order terms to facilitate the interpretation of the  morphic coefficients calculated with atomistic simulations. In this appendix, we write the explicit form of the internal energy in the Lagrangian description, as defined by Rouchy et al. in Ref.\cite{Rouchy1979}. The internal energy it is expanded in series of the Lagrangian strain tensor $\eta_{ij}$ which is defined as\cite{Rouchy1979}
\begin{equation}
\begin{aligned}
    \eta_{ij}=\epsilon_{ij}+\frac{1}{2}\sum_{k}\left(\epsilon_{ki}+\omega_{ki}\right)\left(\epsilon_{kj}+\omega_{kj}\right),
    \label{eq:eta}
\end{aligned}
\end{equation}
where $\epsilon_{ij}$ is given by Eq.\ref{eq:disp_vec} and
\begin{equation}
\begin{aligned}
     \omega_{ij}=\frac{1}{2}\left(\frac{\partial u_{i}}{\partial r_{j}}-\frac{\partial u_j}{\partial r_i}\right),\quad\quad i,j=x,y,z
    \label{eq:omega}
\end{aligned}
\end{equation}
The use of the Lagrangian tensor (finite strain theory) is required to theoretically describe some MEL effects on sound velocity, like the rotational-magnetostrictive effect (function $R(\lambda)$ in Eq.\ref{eq:dfani})\cite{Rouchy1979,ROUCHY198069,booktremolet}. For the analysis of the results given by the atomistic simulations in this work, it may be enough to consider the following terms in the internal energy per volume\cite{Rouchy1979}
\begin{equation}
\begin{aligned}
     E=E_{el}^I+E_{el}^{II}+E_{me}^{I}+E_{me}^{II}+E_{a},
    \label{eq:E_int_1}
\end{aligned}
\end{equation}
where $E_{el}^I$ and $E_{el}^{II}$ are the elastic energy terms up to second and third order in the  Lagrangian strain, respectively, $E_{me}^I$ and $E_{me}^{II}$ are the magentoelastic energy terms up to first and second order in the  Lagrangian strain, respectively, while $E_a$ is the unstrained MCA energy. These terms are given by (cubic crystals, point groups $432$, $\bar{4}3m$, $m\bar{3}m$)\cite{Rouchy1979}
\begin{equation}
\begin{aligned}
     E_{el}^I & = \frac{C_{11}}{2}(\eta_{xx}^2+\eta_{yy}^2+\eta_{zz}^2)+C_{12}(\eta_{xx}\eta_{yy}+\eta_{xx}\eta_{zz}+\eta_{yy}\eta_{zz})\\
& + 2C_{44}(\eta_{xy}^2+\eta_{yz}^2+\eta_{zx}^2) ,\\
     E_{el}^{II} & =  \frac{\tilde{C}_{112}}{2}(\eta_{xx}^2[\eta_{yy}+\eta_{zz}]+\eta_{yy}^2[\eta_{xx}+\eta_{zz}]+\eta_{zz}^2[\eta_{yy}+\eta_{xx}])\\
     & + \tilde{C}_{123}\eta_{xx}\eta_{yy}\eta_{zz}+2\tilde{C}_{144}(\eta_{xx}\eta_{yz}^2+\eta_{yy}\eta_{zx}^2+\eta_{zz}\eta_{xy}^2)\\
     & + 2\tilde{C}_{155}(\eta_{yz}^2[\eta_{yy}+\eta_{zz}]+\eta_{zx}^2[\eta_{zz}+\eta_{xx}]+\eta_{xy}^2[\eta_{xx}+\eta_{yy}])\\
     & + 8\tilde{C}_{456}\eta_{xy}\eta_{yz}\eta_{zx},\\
     E_{me}^{I} & = B^{\alpha,2}U^{\alpha}\tilde{K}^{\alpha,2}+B^{\gamma,2}[U^{\gamma}_1\tilde{K}_1^{\gamma,2}+U^{\gamma}_2\tilde{K}_2^{\gamma,2}]\\
     & + B^{\epsilon,2}[U^{\epsilon}_1\tilde{K}_1^{\epsilon,2}+U^{\epsilon}_2\tilde{K}_2^{\epsilon,2}+U^{\epsilon}_3\tilde{K}_3^{\epsilon,2}]\\
     & =\frac{1}{3}B^{\alpha,2}(\eta_{xx}+\eta_{yy}+\eta_{zz})\\
     & + B^{\gamma,2}\left(\left[\alpha_x^2-\frac{1}{3}\right]\eta_{xx}+\left[\alpha_y^2-\frac{1}{3}\right]\eta_{yy}+\left[\alpha_z^2-\frac{1}{3}\right]\eta_{zz}\right)\\
     & + 2 B^{\epsilon,2}(\alpha_{y}\alpha_{z}\eta_{yz}+\alpha_{z}\alpha_{x}\eta_{zx}+\alpha_{x}\alpha_{y}\eta_{xy}),\\
     E_{me}^{II} & = \tilde{M}_1^{\alpha,2}\Pi_1^{\alpha}\tilde{K}^{\alpha,2}+\tilde{M}_2^{\alpha,2}\Pi_2^{\alpha}\tilde{K}^{\alpha,2}+\tilde{M}_3^{\alpha,2}\Pi_3^{\alpha}\tilde{K}^{\alpha,2}\\
     & + \tilde{M}_1^{\gamma,2}[\Pi^{\gamma}_{1,1}\tilde{K}_1^{\gamma,2}+\Pi^{\gamma}_{1,2}\tilde{K}_2^{\gamma,2}] + \tilde{M}_2^{\gamma,2}[\Pi^{\gamma}_{2,1}\tilde{K}_1^{\gamma,2}+\Pi^{\gamma}_{2,2}\tilde{K}_2^{\gamma,2}]\\
     & + \tilde{M}_3^{\gamma,2}[\Pi^{\gamma}_{3,1}\tilde{K}_1^{\gamma,2}+\Pi^{\gamma}_{3,2}\tilde{K}_2^{\gamma,2}]\\
     & + \tilde{M}_1^{\epsilon,2}[\Pi^{\epsilon}_{1,1}\tilde{K}_1^{\epsilon,2}+\Pi^{\epsilon}_{1,2}\tilde{K}_2^{\epsilon,2}+\Pi^{\epsilon}_{1,3}\tilde{K}_3^{\epsilon,2}]\\
     & +\tilde{M}_2^{\epsilon,2}[\Pi^{\epsilon}_{2,1}\tilde{K}_1^{\epsilon,2}+\Pi^{\epsilon}_{2,2}\tilde{K}_2^{\epsilon,2}+\Pi^{\epsilon}_{2,3}\tilde{K}_3^{\epsilon,2}]\\
     & + \tilde{M}_3^{\epsilon,2}[\Pi^{\epsilon}_{3,1}\tilde{K}_1^{\epsilon,2}+\Pi^{\epsilon}_{3,2}\tilde{K}_2^{\epsilon,2}+\Pi^{\epsilon}_{3,3}\tilde{K}_3^{\epsilon,2}],\\
     E_{a} & = \tilde{K}^{\alpha,0}\tilde{K}^{\alpha,2}\tilde{V}^2+\tilde{K}^{\alpha,0}\tilde{K}^{\alpha,4}\tilde{V}^4,
    \label{eq:E_int_2}
\end{aligned}
\end{equation}
where $C_{ij}$ and $\tilde{C}_{ijk}$ are the second and third order in the Lagrangian strain elastic constants, respectively, $B^{\mu}$ and $\tilde{M}_{i}^{\mu,2}$ ($\mu=\alpha,\beta,\epsilon$) are the first and second order in the Lagrangian strain  MEL constants (up to second order in the direction cosine of magnetization $\boldsymbol{\alpha}$), respectively, while $\tilde{V}^i$ are the MCA constants. The quantities $\tilde{K}^{\mu}_i$ are the cubic harmonic polynomials in terms of $\boldsymbol{\alpha}$, that is\cite{Rouchy1979}
\begin{equation}
\begin{aligned}
     \tilde{K}^{\alpha,0}(\boldsymbol{\alpha}) & = \sqrt{3},\quad \tilde{K}^{\alpha,2}(\boldsymbol{\alpha})  = \frac{1}{\sqrt{3}},\\
     \tilde{K}^{\alpha,4}(\boldsymbol{\alpha}) & = \frac{1}{\sqrt{3}}\left(\alpha_x^4+\alpha_y^4+\alpha_z^4-\frac{3}{5}\right),\\
     \tilde{K}^{\gamma,2}_1(\boldsymbol{\alpha}) & = \sqrt{\frac{2}{3}}\left(\alpha_z^2-\frac{\alpha_x^2+\alpha_y^2}{2}\right), \: \tilde{K}^{\gamma,2}_2 (\boldsymbol{\alpha}) = \frac{1}{\sqrt{2}}\left(\alpha_x^2-\alpha_y^2\right),\\
     \tilde{K}^{\epsilon,2}_1 (\boldsymbol{\alpha}) & = \sqrt{2}\alpha_y\alpha_z, \: \tilde{K}^{\epsilon,2}_2(\boldsymbol{\alpha})  = \sqrt{2}\alpha_z\alpha_x, \: \tilde{K}^{\epsilon,2}_3 (\boldsymbol{\alpha}) = \sqrt{2}\alpha_x\alpha_y.
    \label{eq:K}
\end{aligned}
\end{equation}
The quantities $U^{\mu}_{i}$ are linear strain operators for cubic crystals\cite{Rouchy1979}
\begin{equation}
\begin{aligned}
     U^{\alpha} & = \frac{1}{\sqrt{3}}\left(\eta_{xx}+\eta_{yy}+\eta_{zz}\right),\\
     U^{\gamma}_{1} & = \sqrt{\frac{2}{3}}\left(\eta_{zz}-\frac{\eta_{xx}+\eta_{yy}}{2}\right),\quad  U^{\gamma}_{2}  = \frac{1}{\sqrt{2}}\left(\eta_{xx}-\eta_{yy}\right)\\
      U^{\epsilon}_{1} & = \sqrt{2}\eta_{yz},\quad U^{\epsilon}_{2}  = \sqrt{2}\eta_{zx}, \quad U^{\epsilon}_{3}  = \sqrt{2}\eta_{xy}, 
    \label{eq:U}
\end{aligned}
\end{equation}

while $\Pi^{\mu}_{i,j}$ are quadratic strain operators for cubic crystals\cite{Rouchy1979}

\begin{equation}
\begin{aligned}
     \Pi^{\alpha}_1 & = \frac{1}{\sqrt{3}}\left(\eta_{xx}^2+\eta_{yy}^2+\eta_{zz}^2\right),\\
     \Pi^{\alpha}_2 & = \frac{1}{\sqrt{3}}\left(\eta_{xx}\eta_{yy}+\eta_{yy}\eta_{zz}+\eta_{xx}\eta_{zz}\right), \\
     \Pi^{\alpha}_3 & = \frac{1}{\sqrt{3}}\left(\eta_{yz}^2+\eta_{zx}^2+\eta_{xy}^2\right),\\
     \Pi^{\gamma}_{1,1} & = \sqrt{\frac{2}{3}}\left(\eta_{zz}^2-\frac{\eta_{xx}^2+\eta_{yy}^2}{2}\right),\quad  \Pi^{\gamma}_{1,2}  = \frac{1}{\sqrt{2}}\left(\eta_{xx}^2-\eta_{yy}^2\right)\\
     \Pi^{\gamma}_{2,1} & = \sqrt{\frac{2}{3}}\left(\eta_{xx}\eta_{yy}-\frac{\eta_{yy}\eta_{zz}+\eta_{zz}\eta_{xx}}{2}\right),\\
       \Pi^{\gamma}_{2,2} & = \frac{1}{\sqrt{2}}\left(\eta_{yy}\eta_{zz}-\eta_{zz}\eta_{xx}\right),\\
      \Pi^{\gamma}_{3,1} & = \sqrt{\frac{2}{3}}\left(\eta_{xy}^2-\frac{\eta_{yz}^2+\eta_{zx}^2}{2}\right),\quad  \Pi^{\gamma}_{3,2}  = \frac{1}{\sqrt{2}}\left(\eta_{yz}^2-\eta_{zx}^2\right),\\ 
      \Pi^{\epsilon}_{1,1} & = \sqrt{2}\eta_{xx}\eta_{yz},\quad \Pi^{\epsilon}_{1,2}  = \sqrt{2}\eta_{yy}\eta_{zx}, \quad \Pi^{\epsilon}_{1,3}  = \sqrt{2}\eta_{zz}\eta_{xy}, \\ 
      \Pi^{\epsilon}_{2,1} & = \sqrt{2}(\eta_{yy}+\eta_{zz})\eta_{yz},\quad \Pi^{\epsilon}_{2,2}  = \sqrt{2}(\eta_{zz}+\eta_{xx})\eta_{zx},\\
       \Pi^{\epsilon}_{2,3} & = \sqrt{2}(\eta_{xx}+\eta_{yy})\eta_{xy}, \\ 
      \Pi^{\epsilon}_{3,1} & = \sqrt{2}\eta_{zx}\eta_{xy},\quad \Pi^{\epsilon}_{3,2}  = \sqrt{2}\eta_{xy}\eta_{yz}, \quad \Pi^{\epsilon}_{3,3}  = \sqrt{2}\eta_{yz}\eta_{zx}. 
    \label{eq:Pi}
\end{aligned}
\end{equation}
The fractional change in length can be obtained from the minimization of the elastic and MEL energy\cite{CLARK1980531}. It can also be written as an expansion of cubic harmonic polynomials with respect to $\boldsymbol{\alpha}$ and measuring length direction $\boldsymbol{\beta}$, that is\cite{Rouchy1979}
\begin{equation}
\begin{aligned}
    & \frac{l-l_0}{l_0}\Bigg\vert_{\boldsymbol{\beta}}^{\boldsymbol{\alpha}} = \lambda^{\alpha,2} \tilde{K}^{\alpha,2}(\boldsymbol{\alpha})\tilde{K}^{\alpha,2}(\boldsymbol{\beta})\\
     & + \lambda^{\gamma,2}[\tilde{K}_1^{\gamma,2}(\boldsymbol{\alpha})\tilde{K}_1^{\gamma,2}(\boldsymbol{\beta})+\tilde{K}_2^{\gamma,2}(\boldsymbol{\alpha})\tilde{K}_2^{\gamma,2}(\boldsymbol{\beta})]\\
    & + \lambda^{\epsilon,2}[\tilde{K}_1^{\epsilon,2}(\boldsymbol{\alpha})\tilde{K}_1^{\epsilon,2}(\boldsymbol{\beta})+\tilde{K}_2^{\epsilon,2}(\boldsymbol{\alpha})\tilde{K}_2^{\epsilon,2}(\boldsymbol{\beta})+\tilde{K}_3^{\epsilon,2}(\boldsymbol{\alpha})\tilde{K}_3^{\epsilon,2}(\boldsymbol{\beta})]\\ 
     & =\frac{1}{3}\lambda^{\alpha,2}+\lambda^{\gamma,2}\left(\alpha_x^2\beta_{x}^2+\alpha_y^2\beta_{y}^2+\alpha_z^2\beta_{z}^2-\frac{1}{3}\right)\\
     & + 2\lambda^{\epsilon,2}(\alpha_x\alpha_y\beta_{x}\beta_{y}+\alpha_y\alpha_z\beta_{y}\beta_{z}+\alpha_x\alpha_z\beta_{x}\beta_{z}).
    \label{eq:delta_l_cub_I}
\end{aligned}
\end{equation}

where 

\begin{equation}
    \begin{aligned}
        \lambda^{\alpha,2} & = -\frac{B^{\alpha,2}}{C_{11}+2C_{12}},\\
        \lambda^{\gamma,2} & = -\frac{B^{\gamma,2}}{C_{11}-C_{12}},\\
        \lambda^{\epsilon,2} & = -\frac{B^{\epsilon,2}}{C_{44}}.
    \label{eq:lamb_cub}
    \end{aligned}
\end{equation}

The morphic coefficients $m^{\mu,2}_i$ are linear combinations of $\tilde{C}_{ijk}$ and $\tilde{M}_{i}^{\mu,2}$, and are given by\cite{Rouchy1979}

\begin{equation}
    \begin{aligned}
        m^{\alpha,2}_1 & = \tilde{M}_1^{\alpha,2} + \frac{1}{2}(\tilde{C}_{111}+2\tilde{C}_{112})\lambda^{\alpha,2},\\
         m^{\alpha,2}_3 & = \tilde{M}_2^{\alpha,2} + (\tilde{C}_{123}+2\tilde{C}_{112})\lambda^{\alpha,2},\\
          m^{\alpha,2}_3 & = \tilde{M}_3^{\alpha,2} + 2(\tilde{C}_{144}+2\tilde{C}_{155})\lambda^{\alpha,2},\\
           m^{\gamma,2}_1 & = \tilde{M}_1^{\gamma,2} + \frac{1}{2}(\tilde{C}_{111}-\tilde{C}_{112})\lambda^{\gamma,2},\\
         m^{\gamma,2}_3 & = \tilde{M}_2^{\gamma,2} + (\tilde{C}_{123}-\tilde{C}_{112})\lambda^{\gamma,2},\\
          m^{\gamma,2}_3 & = \tilde{M}_3^{\gamma,2} + 2(\tilde{C}_{144}-\tilde{C}_{155})\lambda^{\gamma,2},\\
          m^{\epsilon,2}_1 & = \tilde{M}_1^{\epsilon,2} + 2\tilde{C}_{144}\lambda^{\epsilon,2},\\
         m^{\epsilon,2}_3 & = \tilde{M}_2^{\epsilon,2} + 2\tilde{C}_{155}\lambda^{\epsilon,2},\\
          m^{\epsilon,2}_3 & = \tilde{M}_3^{\epsilon,2} + 4\tilde{C}_{456}\lambda^{\epsilon,2}.\\
    \label{eq:morphic_coeff}
    \end{aligned}
\end{equation}
The definitions for $m_i^{\mu,2}$ and $\tilde{M}_i^{\mu,2}$  are the same as in Ref.\cite{Rouchy1979}, but note that they are different to the expressions used in Ref.\cite{booktremolet}. The MEL constants $B^{\mu,2}$, magnetostrictive coefficients $\lambda^{\mu,2}$ and MCA constant $\tilde{V}^4$ are related to the corresponding properties of the spin-lattice model given in Table \ref{table:data_properties} through\cite{nieves2021spinlattice_prb}
\begin{equation}
    \begin{aligned}
        \lambda^{\gamma,2} & = \frac{3}{2}\lambda_{001}, \quad
        \lambda^{\epsilon,2}  = \frac{3}{2}\lambda_{111},\\
        B^{\gamma,2} & =b_1,\quad  B^{\epsilon,2}=b_2, \quad \tilde{V}^4=-\frac{K_1}{2}.
    \label{eq:lamb_cub_sl}
    \end{aligned}
\end{equation}

One final remark about the definition of the MEL constants. By restricting the following analysis to the infinitesimal strain theory, where the Lagrangian tensor $\eta_{ij}$ is replaced by the strain tensor $\epsilon_{ij}$, the MEL energy $E_{me}^I$ in Eq.\ref{eq:E_int_2} becomes
\begin{equation}
    \begin{aligned}
        E_{me}^{I} & = \frac{1}{3}B^{\alpha,2}(\epsilon_{xx}+\epsilon_{yy}+\epsilon_{zz})\\
     & + B^{\gamma,2}\left(\left[\alpha_x^2-\frac{1}{3}\right]\epsilon_{xx}+\left[\alpha_y^2-\frac{1}{3}\right]\epsilon_{yy}+\left[\alpha_z^2-\frac{1}{3}\right]\epsilon_{zz}\right)\\
     & + 2 B^{\epsilon,2}(\alpha_{y}\alpha_{z}\epsilon_{yz}+\alpha_{z}\alpha_{x}\epsilon_{zx}+\alpha_{x}\alpha_{y}\epsilon_{xy}).
    \label{eq:Eme_I1}
    \end{aligned}
\end{equation}
This term is typically rewritten using other definition of the MEL constants as\cite{CLARK1980531}
\begin{equation}
    \begin{aligned}
        E_{me}^{I} & =    b_0(\epsilon_{xx}+\epsilon_{yy}+\epsilon_{zz})+b_1(\alpha_x^2\epsilon_{xx}+\alpha_y^2\epsilon_{yy}+\alpha_z^2\epsilon_{zz})\\
    & +  2b_2(\alpha_x\alpha_y\epsilon_{xy}+\alpha_x\alpha_z\epsilon_{xz}+\alpha_y\alpha_z\epsilon_{yz}), 
    \label{eq:Eme_I2}
    \end{aligned}
\end{equation}
where
\begin{equation}
    \begin{aligned}
        b_0 =\frac{1}{3}(B^{\alpha,2}-B^{\gamma,2}), \quad B^{\alpha,2} =3b_0+b_1.
    \label{eq:b0}
    \end{aligned}
\end{equation}
From a mathematical point of view, both forms of the MEL energy are equivalent. However, the definition of MEL constants in Eq.\ref{eq:Eme_I1} has the advantage that fully decouples the isotropic and anisotropic magnetic interactions, which might be helpful in a systematic theoretical analysis of the effects of magnetic interactions on MEL phenomena. Namely, $B^{\alpha,2}$ contains all contribution to the MEL energy from isotropic magnetic interactions like the isotropic exchange, while $B^{\gamma,2}$ and $B^{\epsilon,2}$ contain all contribution to the MEL energy provided by anisotropic magnetic interactions like SOC and crystal
field interactions\cite{booktremolet}. This means that $b_0$  has  contributions from both the isotropic and  anisotropic magnetic interactions, as shown by Eq.\ref{eq:b0}. For example, we can also see this fact in the relationship between volume magnetostriction (induced by the isotropic exchange) and MEL constants
\begin{equation}
    \begin{aligned}
        \omega_s\simeq \lambda^{\alpha,2}=-\frac{B^{\alpha,2}}{C_{11}+2C_{12}}=-\frac{3b_0+b_1}{C_{11}+2C_{12}},
    \label{eq:ws}
    \end{aligned}
\end{equation}
where $b_0$ cannot account for the entire contribution of isotropic exchange interaction to $\omega_s$ by itself.

\bibliography{mybibfile.bib}
 

\end{document}